\documentclass[prd,superscriptaddress,showpacs,preprintnumbers,nofootinbib]{revtex4}
\usepackage{epsfig}
\usepackage{latexsym}
\usepackage{graphicx}
\usepackage{amsmath}
\usepackage{amsfonts}
\usepackage{amssymb}
\usepackage{float}
\usepackage{bm}
\usepackage{ulem}

\newcommand{\mathsym}[1]{{}}

\newcommand{\baz}{\begin{array}{cc}}
\newcommand{\bad}{\begin{array}{ccc}}
\newcommand{\bi}{\begin{itemize}}
\newcommand{\ei}{\end{itemize}}
\newcommand{\ba}{\begin{array}{c}}
\newcommand{\ea}{\end{array}}

\newcommand{\beqa}{\begin{eqnarray}} 
\newcommand{\eeqa}{\end{eqnarray}}

\def\be{\begin{equation}}
\def\ee{\end{equation}}
\def\mET{E_T \hspace{-.9em}/\;\:}
\newcommand{\bea}{\begin{equation} \begin{array}{c}}
\newcommand{\eea}{ \end{array} \end{equation}}

\def\gs{\mathrel{
   \rlap{\raise 0.511ex \hbox{$>$}}{\lower 0.511ex \hbox{$\sim$}}}}
\def\ls{\mathrel{
   \rlap{\raise 0.511ex \hbox{$<$}}{\lower 0.511ex \hbox{$\sim$}}}}

\begin{document}

\title{Vacuum Stability Constraints on the Minimal Singlet TeV Seesaw Model}

\author{Subrata Khan}
\email{subrata@prl.res.in}
\affiliation{Physical Research Laboratory, Navrangpura, Ahmedabad 380 009, India}

\author{Srubabati Goswami}
\email{sruba@prl.res.in}
\affiliation{Physical Research Laboratory, Navrangpura, Ahmedabad 380 009, India}

\author{Sourov Roy}
\email{tpsr@iacs.res.in}
\affiliation{Department of Theoretical Physics, Indian Association for the Cultivation of Science,
2A $\&$ 2B Raja S.C. Mullick Road, Kolkata 700 032, India}

\date{\today}
\begin{abstract}
\noindent
We consider the minimal seesaw model in which two gauge singlet right handed neutrinos
with opposite lepton numbers are added to the Standard Model. In this model, the smallness 
of the neutrino mass is explained by the tiny lepton number violating coupling  between 
one of the  singlets with the standard left-handed neutrinos. This allows one to have
the right handed neutrino  mass at the TeV scale as well as appreciable  mixing between 
the light and heavy states. This model is  fully reconstructible in terms of the neutrino
oscillation parameters apart from the overall coupling strengths. We show that the overall 
coupling strength $y_\nu$ for the  Dirac type coupling between the left handed neutrino and 
one of the singlets can be restricted by consideration of the (meta)stability bounds on the 
electroweak vacuum. 
In this scenario the lepton flavor violating decays of charged leptons can be appreciable
which can put further constraint on $y_\nu$,  for right-handed neutrinos at TeV scale.
We  discuss the combined constraints on $y_\nu$ for this scenario from the process 
$\mu \rightarrow e \gamma$ and from the consideration of vacuum (meta)stability constraints 
on the Higgs self coupling. We also briefly discuss the implications  for neutrinoless double
beta decay and possible signatures of the model that can be expected at colliders.
\end{abstract}



\maketitle
\section{Introduction} 

The ATLAS and CMS collaboration of LHC experiment has reported 
the observation of a new neutral boson. The mass of this particle 
is reported to be \cite{cms,atlas}
\begin{equation} 
M=  126.0 \pm 0.4 \pm 0.4 ~{\mathrm {GeV}}~~ ({\mathrm {ATLAS}});
~~~~M = 125.3 \pm 0.4 \pm 0.5 ~{\mathrm {GeV}}~~ ({\mathrm {CMS}}).
\label{hm} 
\end{equation} 
This is now known to be the Standard Model Higgs boson which has eluded 
scientists so far. Further analysis and data would confirm this and would also 
explore if there is any hint of new physics beyond the Standard Model.   

The Higgs boson is responsible for giving mass to all the fundamental 
particles. However how neutrinos get their mass still remains an enigma.  
The existence of neutrino masses and flavor mixing are already established 
by the oscillation experiments. The oscillation of three known neutrinos are 
characterized by two mass squared differences with values $\sim 10^{-4}$ eV$^2$ 
and $\sim 10^{-3}$ eV$^2$. There is also a mass bound on the sum-total of the 
light neutrino masses from cosmology  which has become more stringent after Planck 
results $\sum{m_i} < 0.23$ eV  \cite{Ade:2013lta}. The mass squared differences 
inferred from oscillation data along with the cosmological mass bound indicate 
that the neutrino masses are much smaller than the corresponding charged fermion 
masses. 

The most natural mechanism which can explain such small masses is the 
Seesaw Mechanism \cite{Minkowski:1977sc,Yanagida:1980,Gell-Mann:1980vs,Glashow:1979vf,Mohapatra:1979ia,Schechter:1980gr} 
which postulates a heavy particle at some high scale determined by the mass of this particle. 
Usually this scale is $M \sim 10^{14}$ GeV to account for the smallness of the neutrino mass.   

Since the LHC started operation, a natural question which has been explored in the literature 
quite extensively is the possibility of observing signature of seesaw at the LHC. This will require 
the mass of the heavy particles to be of the order of TeV scale. For the canonical type-I seesaw mechanism
this is difficult and one has to appeal to cancellations coming from flavor 
symmetries \cite{Adhikari:2010yt,Kersten:2007vk,Pilaftsis:1991ug}. Another option to relate small 
neutrino mass to TeV scale physics is  provided by the inverse or linear seesaw models
\cite{valle-mohapatra,Gu:2010xc,Zhang:2009ac,Hirsch:2009mx}. 
Such models contain additional singlet states. In these models smallness of 
neutrino mass can be explained by the small lepton number violation in the 
couplings (dimensionless and/or dimensionful) of the singlet fields
and the scale of new physics can be at TeV in a natural way. 

Such models can have a number of phenomenological consequences.
Non-unitary mixing between light and heavy particles can be large 
in these models and can be probed at colliders\cite{Keung:1983uu,Bandyopadhyay:2012px,Dev:2012zg,Das:2012ze}. 
Future neutrino factories may also be sensitive to non-unitarity  
\cite{Antusch:2009pm,Goswami:2008mi,Fernandez-Martinez:2007ms}.  
Lepton flavor violating processes can be appreciable \cite{Forero:2011pc,Ilakovac:1994kj}. 
The non-unitary effect can also play a non-trivial role in relating
CP violation responsible for leptogenesis with low energy CP violation 
\cite{Antusch:2009gn,Rodejohann:2009cq}. 

However, assuming the particle observed in CMS and ATLAS is the 
Higgs Boson and its mass to be as given in Eq.(\ref{hm}) opens 
up an avenue for constraining the Dirac Yukawa couplings in 
seesaw models from the consideration of stability of the 
electroweak vacuum \cite{vac-stabnu-ibara,vac-stabnu-earlier}. 
It is well known that because of quantum corrections, the 
Higgs self-coupling $\lambda$ diverges for higher values of Higgs mass 
and goes to negative for low values of Higgs mass near Planck-scale 
($M_{pl}=1.2\times 10^{19}$ GeV). Assuming no new physics between SM and 
the Planck scale, Higgs mass was found to be in the range $126-171$ GeV
for $\lambda$(at $M_{pl}$) to be in the range [$0, \pi$] 
\cite{shapos,lindner,bezrukov,Alekhin:2012py}.
The upper bound called the ``triviality bound"  essentially embodies the
perturbativity of the theory. The lower bound  known as the  ``vacuum stability 
bound" is obtained from the fact that a negative $\lambda$  makes the potential 
unbounded from below and the vacuum would be unstable. In view of the present 
experimental mass range of the Higgs boson \cite{cms,atlas}, it is likely that 
SM vacuum is  metastable \cite{Degrassi:2012ry}. 
The metastability condition implies that the probability of quantum tunnelling 
is small so that the life time 
of the SM vacuum is greater than the age of the Universe.
While this allows 
$\lambda$ to assume  negative values, it cannot be too large
\cite{Isidori:2001bm,Espinosa:2007qp}.
This in  turn implies 
the mass range of the Higgs boson as $105-126$ GeV 
\cite{Ellis:2009tp,EliasMiro:2011aa,Degrassi:2012ry}.

The presence of new Yukawa couplings in seesaw models
modifies the $\beta$ function of Higgs self-coupling.  
In the conventional type-I seesaw model, 
generation of small neutrino mass requires 
the mass scale of the singlet to be of the order of $10^{14}$ GeV
for Dirac Yukawa Coupling $Y_{\nu} \sim {\cal{O}}(1)$. 
It was observed in \cite{vac-stabnu-ibara} that 
the presence of this extra coupling increases the 
lower bound of the Higgs mass from vacuum stability constraints, 
gradually  reaching  the perturbativity bound. However as the mass 
scale of the heavy field is lowered, $Y_{\nu}$ has to become less 
in order to get $m_\nu \sim 0.1$ eV and below a certain value of the
mass scale the additional contribution does not play any significant role.
Nevertheless, from the point of view of relevance at LHC many models 
have been considered in the literature which can give rise to small 
neutrino masses with a relatively large Yukawa coupling even with the 
heavy field at the TeV scale. 
Hence in such models the effect of the Yukawa term 
can be significant in the running of $\lambda$. 
Moreover, as the neutrino Yukawa runs from TeV
to Planck scale, the effect can be large \cite{Rodejohann:2012px,Chakrabortty:2012np}. 
Since the presence of this term drives $\lambda$ towards a more negative value, 
constraints were obtained on the Yukawa coupling strength  
from conditions of absolute stability which implies $\lambda(M_{Pl}) \geq 0$
\cite{vac-stabnu-ibara,vac-stabnu-earlier,Rodejohann:2012px,Chakrabortty:2012np,
Chen:2012faa}. 

In this work we consider the Minimal Linear Seesaw Model 
(MLSM) which can naturally accommodate TeV scale singlets.  
We show that it is possible to constrain the unknown Dirac-Yukawa 
coupling strength $y_\nu$ in this model from the considerations of 
vacuum (meta)stability of the scalar potential. Recently, vacuum 
stability bounds on the Dirac Yukawa coupling in TeV scale seesaw model 
have been obtained in \cite{Rodejohann:2012px,Chakrabortty:2012np}. 
However, metastability constraints in the context of
seesaw models including the one loop effect of heavy neutrinos towards 
the effective potential, have not been studied 
in the literature so far. In canonical seesaw models one needs to make 
some assumptions about the structure of the Dirac type Yukawa matrix $Y_\nu$ 
and the right handed Majorana mass matrix $M_R$. On the other hand, for 
MLSM this is already completely determined in terms of oscillation 
parameters apart from the overall Yukawa coupling strengths \cite{gavela-h3} 
and hence one need not make any further assumptions on the structure of 
the mass matrices. This feature makes it particularly suitable for studying 
vacuum (meta)stability constraints. 
Since the heavy singlet states in this 
model are at TeV scale, lepton flavor violating decays of charged leptons are 
not suppressed and from the bound on the branching ratios of these processes 
it is possible to constrain $y_\nu/M_R$. We consider the bound on the process 
$\mu \rightarrow e \gamma$ and discuss the upper bound obtained on $y_\nu$ together 
with the constraints from vacuum (meta)stability as a function of the mass scale $M_R$.
We also comment on the implications of this model for neutrinoless double beta 
decay and discuss the possible collider signatures. 

The plan of the paper is as follows. In the next section we discuss
the minimal singlet seesaw mass matrix. Section III describes the running of 
the self coupling $\lambda$ and investigates the stability and metastability 
of the scalar potential in the context of SM given the current bounds on the 
mass of Higgs, top and the strong coupling constant $\alpha_s$.
We also  obtain the constraints on the Yukawa coupling strength $y_\nu$ 
from vacuum (meta)stability in MLSM for known values of Higgs masses.  
In  section IV we study the phenomenological implications of this model for 
charged  Lepton Flavor Violation (LFV). In section V we consider  neutrinoless 
double beta decay ($0\nu\beta\beta$) in this model. In the next section we comment 
on possible collider signatures  and finally present the  conclusions in section VII.

\section{Singlet Seesaw Models} 

The Yukawa part of the most general Lagrangian involving extra singlet states
can be written as 
\begin{eqnarray}
-{\cal{L}} = \overline{N}_R Y_{\nu} \tilde{\phi}^{\dag} l_{L} 
 +  \overline{S} Y_{S} \tilde{\phi}^{\dag} l_{L}
+  \overline{S}M_R N_R^c + \frac{1}{2}\overline{S} \mu S^c 
+ \frac{1}{2} \overline{N_R} {M_N} N_R^c 
+ {\mathrm h.c.}, 
\label{lag:full}
\end{eqnarray}
where $l_L = (\nu_x,x)_L^T$, $x = {e, \mu, \tau}$.   
$l_L, N_R$ and $S$ have lepton number $1,1,-1$, respectively.
After spontaneous symmetry breaking, the $\phi$ field acquires a vacuum 
expectation value ($v/\sqrt{2}$) and $Y_\nu v/\sqrt{2} = m_D $ 
gives rise to the Dirac mass term
while the term $Y_s v/\sqrt{2} = m_S$ breaks lepton number. 
In the above Lagrangian lepton number violation stems from the
terms with coefficients $Y_s$, $\mu$ and $M_N$  and thus  
the symmetry of the Lagrangian is enhanced (lepton number becomes an exact symmetry)
in the absence of these terms.  
Therefore, these coefficients can be small in a natural way
(i.e. there is no {\it fine tuning} or {\it unnaturalness} in
keeping these terms to be very small) according to 't Hooft's naturalness criterion.

The neutrino mass matrix in the $(\nu_L,N_R^c,S^c)$ basis can be written as 
\be
M_\nu = 
\bad
\begin{pmatrix}
0 & m_D^T &  m_S^T \\
m_D & M_N & M_R^T \\
m_S & M_R & \mu \\
\end{pmatrix}.
\end{array} 
\label{massmatrix} 
\ee
In the literature many variants of this model have  been considered. 
\\
\underline{Inverse Seesaw} 
\\
The conventional inverse seesaw models assume  the terms 
$m_S$ and $M_N$ in Eq. (\ref{massmatrix}) to be zero. 
The model is lepton number conserving in the limit $\mu$ tending to zero.
The minimal inverse seesaw model considered in the literature \cite{malinsky} 
consists of  $3 \nu_L + 2 N_R +2 S$. 
The model with  $ 3\nu_L + 1 N_R + 1 S$ is a 
$5 \times 5$ matrix with rank 3. 
Thus there are two zero eigenvalues 
of this matrix which is not consistent with low energy phenomenology.
The model consisting of $3\nu_L + 2 N_R +1 S$ 
is a $6\times 6$ matrix with rank 5. Thus there is one zero eigenvalue. 
However, this belongs to the $(N_R,S)$ block and hence this scenario is not considered 
if one assumes  that there are no light singlets.
In principle the Majorana mass term of $N_R$ can be included \cite{Hu:2011ac,Kang:2006sn}, although
this does not change the structure of the effective light neutrino mass matrix
at the leading order \cite{Bazzocchi:2010dt,Dev:2012sg}.  
\\
\underline{Linear Seesaw} 
\\
In the so called linear seesaw models \cite{Gu:2010xc,Zhang:2009ac,Hirsch:2009mx} 
one retains  the $\nu-S$ term in the Lagrangian through the
Yukawa coupling matrix  $Y_s$ and makes 
the $\mu$ and the $M_N$ term to be zero.  
In these models  lepton number violation stems from the 
term containing $Y_s$.   
In the limit  $M_R >> m_D, m_S$ the above mass matrix can be diagonalized using the seesaw 
approximation and in the leading order the effective light neutrino mass matrix $m_{light}$ can be expressed as
\begin{eqnarray}
m_{\mathrm light} =  m_D^T {M_R}^{-1} m_S  + m_S^T {M_R}^{-1} m_D. 
\label{eq:mlightlinear}
\end{eqnarray} 
Since this contains only one power of the Dirac mass term it is called 
linear seesaw. 

One can make an order of magnitude estimate of the various terms to check 
the conditions required to get $m_\nu \sim 0.1$ eV. 
Assuming typical values $m_D \sim 100$ GeV  (Yukawa coupling strength 
$Y_\nu \sim {{\cal{O}} (1)}$, $v \sim 100$ GeV) and $M_R = 1$ TeV 
one needs $Y_s \sim 10^{-11}$.  
In the heavy sector we get two degenerate neutrinos of mass $\sim$ TeV. 
The minimal model consists of adding just two singlet states 
$N_R$  and $S$. The rank of the 3+1+1 mass matrix is 4 corresponding to one 
zero mass eigenvalue. The Majorana mass term $M_N$ can also be included
which would lift the degeneracy between the heavy states, 
However, the contribution of this term to the light neutrino mass matrix is
sub-dominant \cite{Bazzocchi:2010dt,Dev:2012sg}.
\\
\underline{Inverse + Linear Seesaw} 
\\
It is also possible to keep both the terms $m_s$ and $\mu$ in the Lagrangian. 
Then in the limit  $M_R >> m_D, m_S$ and in the leading order the effective light 
neutrino mass matrix $m_{light}$ can be expressed as
\begin{eqnarray}
m_{\mathrm light} = - m_D^T \frac{\mu}{M_R^2} m_D + m_D^T \frac{1}{M_R} m_S  + m_S^T \frac{1}{M_R} m_D. 
\end{eqnarray} 
In this case, for $M_R \sim ~ \text{TeV}$, one needs $\mu \sim ~ 10^{-8}$ GeV and $Y_s \sim 10^{-11}$.
This hybrid scenario allows one to reconstruct $Y_\nu$ and the combination 
$Y_s - \frac{\mu}{2 M_R} Y_\nu$. Thus reconstruction of $Y_S$ requires  
another unknown parameter, $\mu$ \cite{gavela-h3}. 
In our subsequent discussion we assume $\mu$ to be zero 
and consider the linear seesaw option. 

\subsection{Minimal Linear Seesaw Model} 

The Minimal Linear Seesaw Model  (MLSM) is defined by the 
mass matrix in Eq. (\ref{massmatrix}) with $M_N\,,\mu=0$ and just 2 singlet states.  
Then the entries $M_R$ are numbers instead of matrices and the 
dimension of the full matrix is $5 \times 5$. 
This can be written as,
\begin{equation}
M_\nu = 
\left(\begin{array}{cc} 
0 & {m_D^\prime}^T \\
{m_D^\prime} &  M  
\end{array}\right),
\label{mrecast}
\end{equation}
where $m_D^{\prime T} = (m_D^T,m_S^T)$.
Now defining M as
\begin{equation}
M = 
\left(\begin{array}{cc} 
0 & M_R \\
M_R & 0 
\end{array}\right),
\label{msinglet} 
\end{equation}
the neutrino mass matrix $M^\nu$ can be diagonalized
by a 5 $\times$ 5 unitary matrix $U_0$ as
\begin{eqnarray}
\label{diagonal}
U_0^T M^\nu U_0 = M_\nu^{\mathrm diag},
\end{eqnarray}
where $M^{\mathrm diag} 
=\mbox{diag}(m_1,m_2,m_3, M_1,M_2)$ 
with mass eigenvalues $m_i$ ($i=1,2,3)$) and $M_j$ ($j=1,2$) for
light and heavy neutrinos respectively.
Following standard procedure of two-step diagonalization
$U_0$ can be expressed as \cite{Grimus:2000vj}
\begin{eqnarray}
U_0= W\, U_{\nu} =
\left(\begin{array}{cc}  
\left(1-\frac{1}{2}\epsilon \right)U & {m_D^{\prime \dagger}} (M^{-1})^{\ast}U_{R}\\
-M^{-1}m_D^{\prime} U & \left(1-\frac{1}{2}\epsilon'\right)U_{R}
\end{array} \right) 
 = \left(\begin{array}{cc}
U_{L} & V\\
S & U_{H }
 \end{array} \right) ,
\label{bdmatrix} 
\end{eqnarray}
where
$W$ is the matrix which brings the
full $5 \times 5$ neutrino matrix,
in the block diagonal form 
\begin{eqnarray}
W^T\begin{pmatrix}
0 & m_{D}^{\prime T}\\
m_{D}^{\prime } & M 
\end{pmatrix}W
=\begin{pmatrix}
m_{\text{light}} & 0\\
0 & m_{\text{heavy}}
\end{pmatrix}.
\label{blockdiagonal}
\end{eqnarray}
$U_\nu = diag(U,U_R)$ diagonalizes the mass matrices in the light and heavy
sector appearing in the upper and lower block of the block diagonal 
matrix respectively. $U_L$ in Eq.(\ref{bdmatrix}) corresponds to $U_{PMNS}$ 
which acquires a non-unitary correction $(1 - \epsilon/2)$.  
The eigenvalues $(M_1,M_2)$ are obtained 
as $(-M_R,M_R)$ corresponding to degenerate neutrinos 
with opposite CP parities. 
The  negative sign in the mass eigenvalues 
can be absorbed in the phases of the diagonalizing matrix $U_R$ giving,   
\begin{equation} 
U_R = \frac{1}{\sqrt{2}}
\begin{pmatrix} 
i & 1 \\ -i & 1 
\end{pmatrix}.
\label{ur} 
\end{equation} 
$\epsilon$ and $\epsilon'$, which characterize the non-unitarity,
are given by
\begin{eqnarray}
\epsilon=m_D^{\prime \dagger} \left(M^{-1}\right)^\ast M^{-1} m_{D}^{\prime},\nonumber \\
\epsilon'=  M^{-1}m_{D}^{\prime}m_D^{\prime \dagger}\left(M^{-1}\right)^\ast.
\label{epsilons} 
\end{eqnarray}
Since Eq.(\ref{mrecast}) is in the standard seesaw form it is straightforward to obtain 
the light neutrino mass matrix 
\begin{eqnarray}
m_{\mathrm light} = m_D^{\prime T} M^{-1} m_D^{\prime}, 
\label{effective-nu-massmatrix}
\end{eqnarray}
in the limit $M_R >> m_D, m_S$. 
Now inserting the expression for $m_D^\prime$, the light neutrino mass matrix    
is the same as that in Eq. (\ref{eq:mlightlinear}).  
Note that the complete mass matrix 
for the minimal model has 7 phases out of which 5 can be rotated away by redefinition of the 
fields. Thus there are 2 independent phases in this matrix. 
We choose the basis in which $M_R$ is real and attach the phases to the elements of 
$Y_\nu$ and $Y_s$. Since $M_\nu$ for this case is of rank 4,
there is one zero eigenvalue. Thus one of the light neutrino states is massless and the two 
remaining masses are completely determined in terms  of the two mass squared differences measured 
in oscillation experiments.

It is very interesting to note that $m_{light}$ for this case is determined in terms of two independent vectors 
\begin{equation} 
Y_\nu  \equiv  y_\nu {\hat{\bf{a}}};~~
Y_S \equiv y_s {\hat{\bf{b}}} 
\end{equation} 
where ${\hat{\bf{a}}}$ and ${\hat{\bf{b}}}$ are complex vectors with unit norm. 
$y_{\nu}$ and $y_{s}$ are the norms of the Yukawa matrices $Y_{\nu}$ and $Y_S$, respectively.
This  feature allows one to completely reconstruct the Yukawa matrices 
$Y_\nu$ and $Y_S$ in terms of the oscillation parameters as \cite{gavela-h3},
\begin{itemize} 
\item
Normal Hierarchy (NH): 
~~~ $(m_1 < m_2 < m_3)$\footnote{The phase factor, $e^{i\frac{\pi}{2}}$, is inserted 
to ensure the positive definiteness of the mass eigen values.}
\begin{eqnarray}
Y_{\nu}&=&\frac{y_{\nu}}{\sqrt{2}}\left(\sqrt{1+\rho}~U_3^{\dag} + e^{i\frac{\pi}{2}}\sqrt{1-\rho}~U_2^{\dag}\right) \nonumber \\
Y_{S}&=&\frac{y_{s}}{\sqrt{2}}\left(\sqrt{1+\rho}~U_3^{\dag} - e^{i\frac{\pi}{2}}\sqrt{1-\rho}~U_2^{\dag}\right)
\label{pmtz-nh} 
\end{eqnarray}
with
\begin{eqnarray}
\rho&=&\frac{\sqrt{1+r}-\sqrt{r}}{\sqrt{1+r}+\sqrt{r}}. 
\end{eqnarray}
$U_i$'s are the columns of the unitary matrix $U$ that diagonalizes
the light neutrino mass matrix ($m_{light}$) above and 
$r$ is the ratio of the solar and atmospheric mass  squared differences
\be
r=\frac{\Delta m_{\odot}^2}{\Delta m^2_{atm}}.
\ee
\item 
Inverted Hierarchy (IH): ~~~$(m_3 << m_2 \approx m_1 )$    
\begin{eqnarray}
Y_{\nu}&=&\frac{y_{\nu}}{\sqrt{2}}\left(\sqrt{1+\rho}~U_2^{\dag} + e^{i\frac{\pi}{2}}\sqrt{1-\rho}~U_1^{\dag}\right) \nonumber \\
Y_{S}&=&\frac{y_{s}}{\sqrt{2}}\left(\sqrt{1+\rho}~U_2^{\dag} - e^{i\frac{\pi}{2}}\sqrt{1-\rho}~U_1^{\dag}\right)
\label{pmtz-ih} 
\end{eqnarray}
with
\be
\rho=\frac{\sqrt{1+r}-1}{\sqrt{1+r}+1}.  
\ee
\end{itemize} 
We use the following form for $U$ 
\begin{equation}
U   =  \left(
 \begin{array}{ccc}
 c_{12} \, c_{13} & s_{12}\, c_{13} & s_{13}\, e^{-i \delta}\\
 -c_{23}\, s_{12}-s_{23}\, s_{13}\, c_{12}\, e^{i \delta} &
 c_{23}\, c_{12}-s_{23}\, s_{13}\, s_{12}\,
e^{i \delta} & s_{23}\, c_{13}\\
 s_{23}\, s_{12}-\, c_{23}\, s_{13}\, c_{12}\, e^{i \delta} &
 -s_{23}\, c_{12}-c_{23}\, s_{13}\, s_{12}\,
e^{i \delta} & c_{23}\, c_{13}
 \end{array}
 \right) P \, ,
\label{upmns_param}
\end{equation}
where 
$c_{ij} = \cos \theta_{ij}$, $s_{ij} = \sin \theta_{ij}$,
$\delta$ is the Dirac CP phase. For the Majorana phase matrix $P$, we use 
$P = {\rm diag}( e^{-i \alpha}, e^{i \alpha},1)$. 
Note that in this case since one of the mass eigenvalues is zero 
there is only one Majorana phase. 
The 3$\sigma$ ranges of the oscillation parameters 
are tabulated in Table \ref{table-osc} \cite{valleosc}. 

\begin{table}[t]
\caption{Present $3\sigma$ range of neutrino
       oscillation parameters. The upper (lower) row corresponds to normal
       (inverted) hierarchy. Values of $\Delta m^2_{21}$ and $\sin^2\theta_{12}$
       are hierarchy independent \cite{valleosc}. \label{table-osc}}
\begin{ruledtabular}
\begin{tabular}{cc}
Parameters & 3$\sigma$ range \\
\hline
$\Delta m^2_{\odot}$ [$10^{-5}$ ${\rm eV}^2$] & 7.12 -- 8.20\\[2.0mm]
\vspace{-1.0mm}$\Delta m^2_{atm}$ [$10^{-3}$ ${\rm eV}^2$] & 2.31 -- 2.74\\
 & 2.21 -- 2.64\\[1.5mm]
$\sin^2{\theta_{12}}$ & 0.27 -- 0.37\\[2.0mm]
\vspace{-1.0mm}$\sin^2{\theta_{23}}$ & 0.36 -- 0.68\\
 & 0.37 -- 0.67\\[1.5mm]
$\sin^2{\theta_{13}}$ & 0.017 -- 0.033\\[1.5mm]
$\delta$ & 0 -- 2$\pi$\\
\end{tabular}
\end{ruledtabular}
\end{table}

From the above forms of the Yukawa matrices it is evident 
that these are completely determined in terms of the 
masses and mixing angles, two unknown phases and the norms of the 
Yukawa couplings $y_\nu$ and $y_s$. 

The minimal Type-I seesaw model also consists of three left-handed and
two gauge singlet right-handed neutrinos \cite{King:1998jw,Raidal:2002xf,Barbieri:2003qd,Ibarra:2003up}. 
However both the right-handed neutrinos are assumed to have the same lepton number. 
Thus both have lepton number conserving Dirac type coupling 
with the light state. In order to have the right-handed neutrinos of this 
model at TeV scale one needs to have small values of the Dirac 
coupling  $Y_\nu \sim 10^{-6}$ unless one allows for 
fine tuning leading to $m_D M_R^{-1} m_D^T = 0$ \cite{Adhikari:2010yt,Kersten:2007vk,Pilaftsis:1991ug}.
Since this coupling is not lepton number violating, its smallness cannot be 
explained naturally.  Also for such small values of the coupling, collider signals would be suppressed  
even though the mass of the right-handed  neutrino is at TeV scale. 

\section{Vacuum stability and Metastability of the Higgs potential}
\subsection{Higgs mass and Vacuum stability and Metastability in SM} 

The tree-level potential of the Higgs field in the Standard Model(SM) 
is given as 
\begin{equation} 
V(\phi) = {\lambda} \left({\phi^\dagger \phi}\right)^2 - m^2 {\phi^\dagger \phi}.
\end{equation} 
This receives quantum corrections from higher order loop diagrams. The physical Higgs mass 
is defined as $m_h^2 = 2\lambda v^2$. Since the Higgs quartic coupling, $\lambda$, receives 
quantum corrections from higher order loop diagrams, it runs with the renormalization scale. 
The Renormalization Group (RG) equation for the Higgs quartic coupling $\lambda$ can be expressed 
in general as 
\begin{equation} 
\mu \frac{d\lambda}{d\mu} = \sum_{i} 
\frac{{\beta_\lambda}^{(i)}}{(16 \pi^2)^i},  
\end{equation} 
where $i$ denotes the $i^{th}$ loop. Assuming SM to be valid up to Planck scale the
$\beta$ function calculated up to $1$ loop is given as, 
\begin{eqnarray} 
\beta_\lambda^{(1)} = 24 \lambda^2 - (\frac{9}{5} g_1^2 + 9 g_2^2) \lambda 
+ \frac{27}{200} g_1^4 + \frac{9}{20} g_1^2 g_2^2 + \frac{9}{8} g_2^2 
+ 4 T \lambda - 2 Y,
\label{eq:betalambda-1loop} 
\end{eqnarray}
where, 
\begin{eqnarray} 
T & = & {\mathrm {Tr}} \left[3 {Y_u}^\dagger Y_u + 3 {Y_d}^\dagger Y_d + {Y_l}^\dagger Y_l \right], \\    
Y & = & {\mathrm{Tr}} \left[3 ({Y_u}^\dagger Y_u)^2 + 3 ({Y_d}^\dagger Y_d)^2 + ({Y_l}^\dagger Y_l)^2\right].  
\end{eqnarray} 
In the above equations, $g_i$ denote the gauge coupling constants 
with $i = 1,2,3$ corresponding to $U(1)$, $SU(2)$ and $SU(3)$ groups respectively.
The above equations include the Grand Unified Theory (GUT) modified 
coupling for the $U(1)$ gauge group. $Y_f$ with $f= u,d,l$ represent the Yukawa coupling matrices for 
the up type, down type quarks and the charged leptons. 
The running behavior is controlled mainly by the top quark mass 
$m_t$ which drives $\lambda$ towards more negative values in the low Higgs mass region. 
The running of the top Yukawa is governed by the following equations 
\begin{eqnarray} 
\beta_{Y_u}^{(1)} = Y_u \left[\frac{3}{2} {Y_u}^\dagger Y_u +
\frac{3}{2} {Y_d}^\dagger Y_d  + T - \left(
\frac{17}{20} g_1^2 + \frac{9}{4} g_2^2 + {8} g_3^2\right)\right].
\label{eq:betayu-1loop} 
\end{eqnarray}
In the numerical work we have used three loop Renormalization Group Equations
(RGE) for $\lambda$, the top Yukawa and the gauge 
couplings\cite{2-loop-rg,Luo:2002ey,Machacek:1983tz,Machacek:1983fi,Machacek:1984zw,Mihaila:2012fm,Chetyrkin:2012rz}. 
Considering the two loop effective potential for the Higgs field, the stability of the electroweak vacuum
at $M_{Pl}$ demands
$\tilde{\lambda} =0$ at $M_{pl}$, where $\tilde{\lambda}$ is the two loop corrected self coupling
defined as \cite{casas_ho,Casas:1996aq}
\begin{eqnarray} 
\tilde{\lambda} &   =  & \lambda
- \frac{1}{32 \pi^2} \left[\frac{3}{8} \left(g_1^2 + g_2^2\right)^2 \left(\frac{1}{3} - 
{\mathrm{log}}\frac{ \left(g_1^2 + g_2^2\right)}{4}\right) + 6 y_t^4 \left({\mathrm{log}} \frac{y_t^2}{2} -1 \right) 
 + \frac{3}{4} g_2^4 
\left(\frac{1}{3} - {\mathrm {log}}\frac{g_2^2}{4}\right) \right] \nonumber \\
& & + \,\frac{Y_t^4}{\left(16\pi^2\right)^2}\left[g_3^2\left\{24\left(\rm{ln}\,\frac{Y_t^2}{2}\right)^2 - 64\,\rm{ln}\,\frac{Y_t^2}{2} + 72 \right\} 
- \frac{3}{2}\,Y_t^2\left\{3\left(\rm{ln}\,\frac{Y_t^2}{2}\right)^2 -16\,\rm{ln}\,\frac{Y_t^2}{2} + 23 + \frac{\pi^2}{3} \right\}\right]. 
\label{lambdatilda}
\end{eqnarray} 
As discussed earlier the constraints from vacuum stability  
and perturbativity limits Higgs mass in the range 126-171 GeV. Therefore 
if the scalar particle observed by the CMS \cite{cms} and ATLAS \cite{atlas} collaboration
is assumed to be the Higgs Boson then the reported mass is 
near the lower bound (upper bound) obtained from vacuum stability (metastability) condition. 
\begin{figure}[ht]
\begin{center}
\includegraphics[width=8.0cm,height=5.5cm]{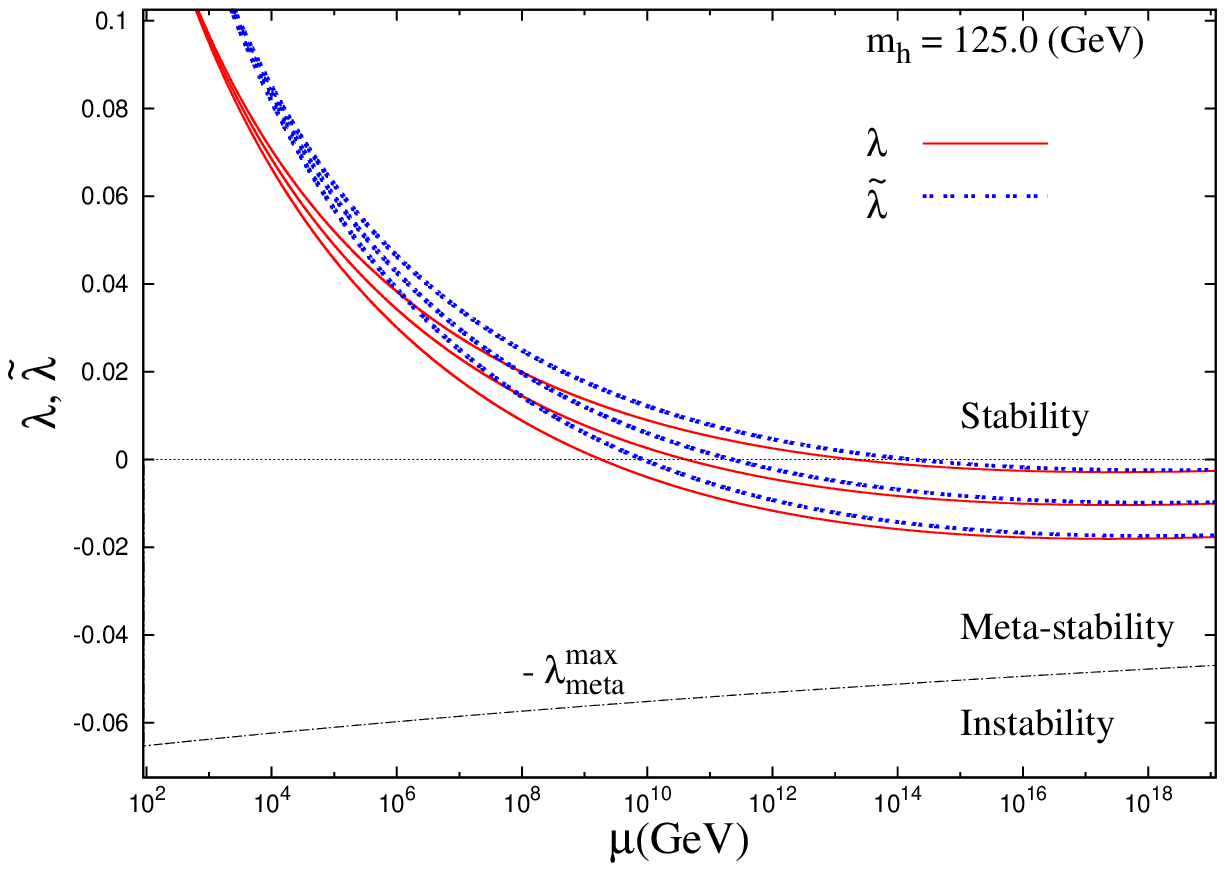}
\includegraphics[width=8.0cm,height=5.5cm]{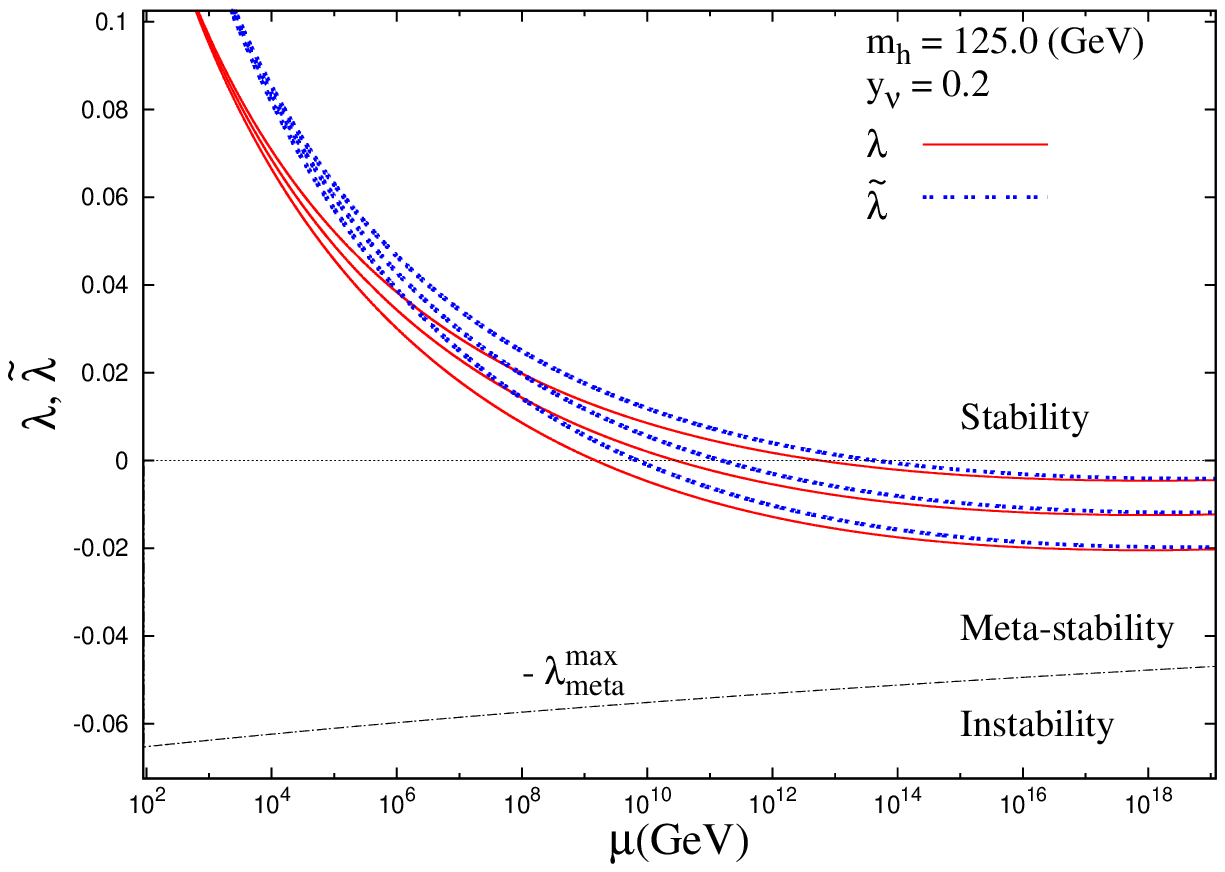}\\
\includegraphics[width=8.0cm,height=5.5cm]{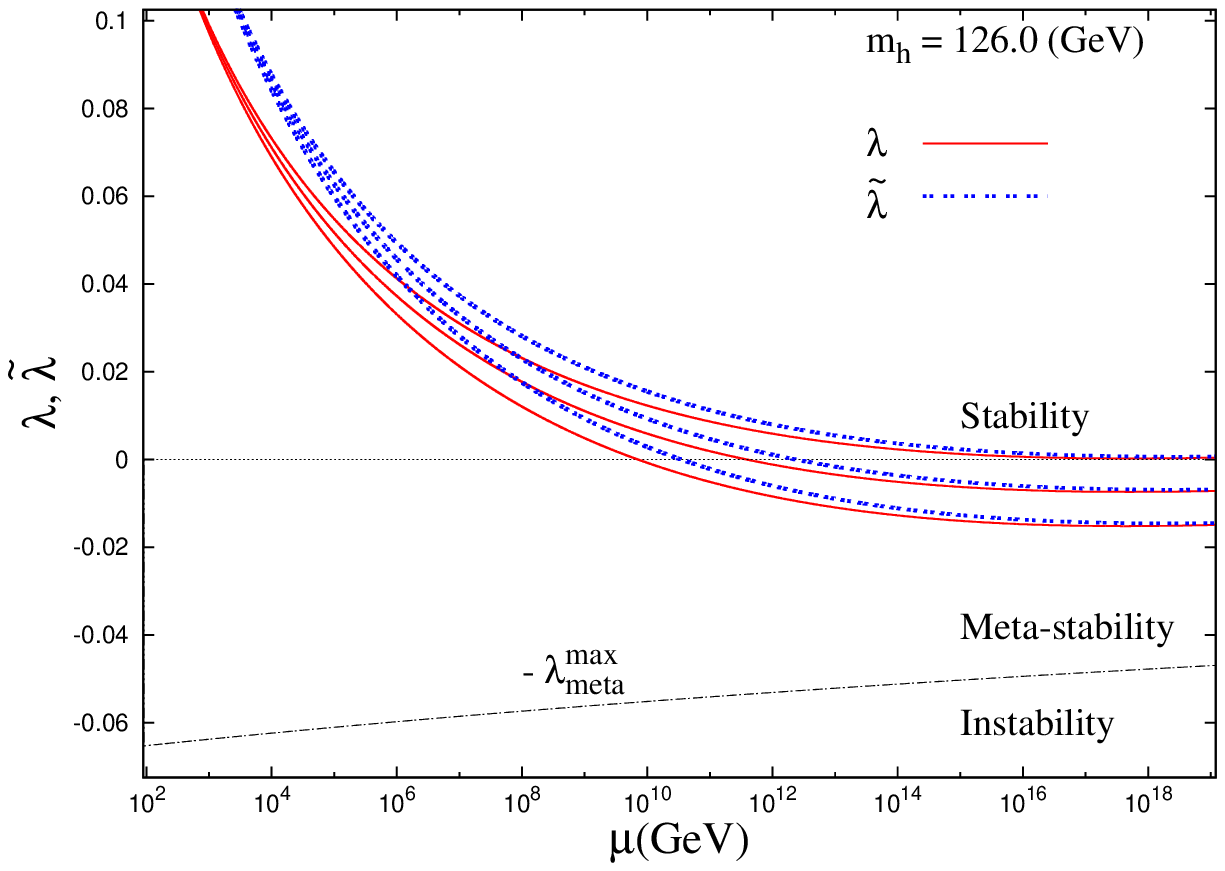}
\includegraphics[width=8.0cm,height=5.5cm]{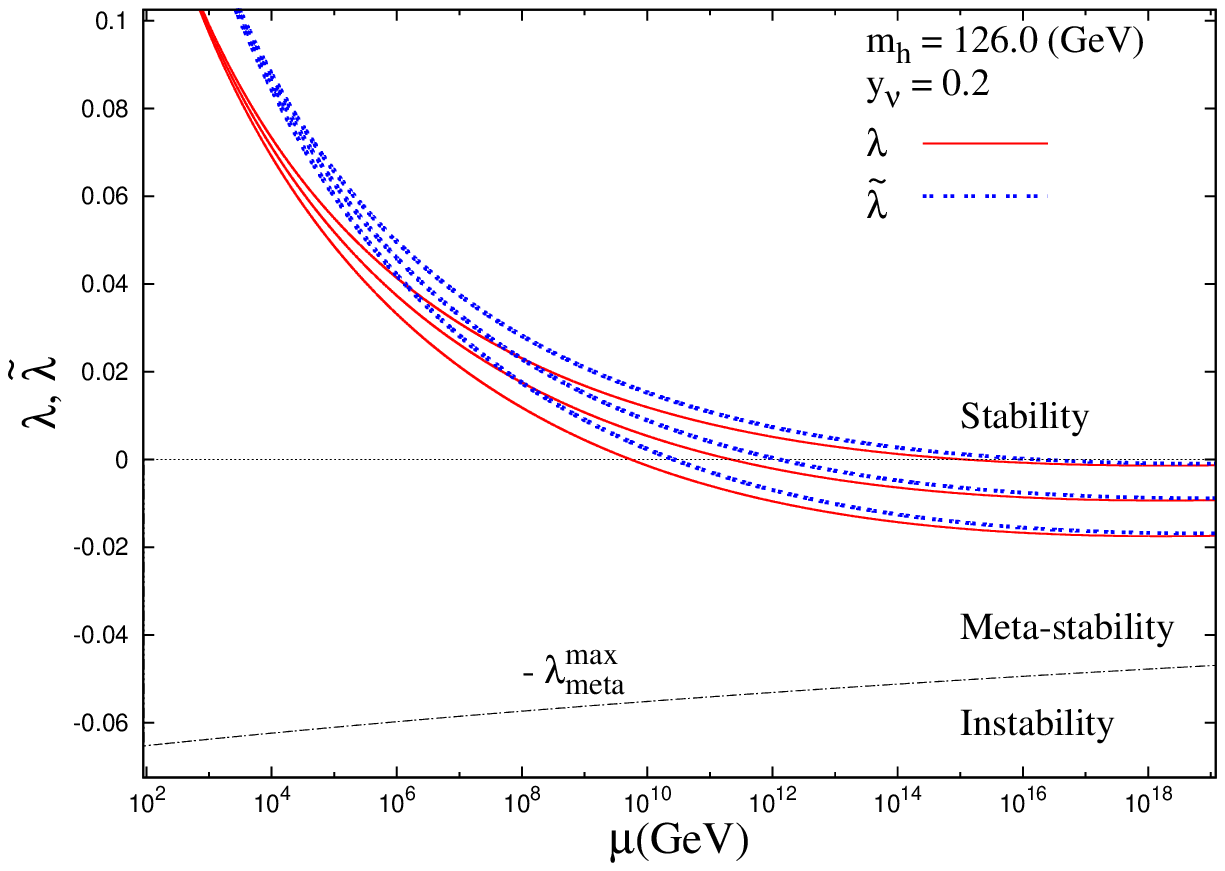} \\
\caption{The left panels show the variation of $\lambda$ and $\tilde{\lambda}$ with the renormalization scale for fixed values 
of the parameters ($m_h\,,m_t\,,\alpha_s$). The upper, middle and lower curves are drawn with the set of parameters 
($m_t,\alpha_s$)=$\{(172.3\,\rm{GeV}\,,0.1191)$, $(173.2\,\rm{GeV}\,,0.1184)$, $(174.1\,\rm{GeV}\,,0.1177)\}$ respectively.
The right panels show the changes after including the Dirac Yukawa coupling parameter $y_\nu$.}
\label{lam-vs-mu}
\end{center}
\end{figure}
In the left panels of Fig. \ref{lam-vs-mu},
we plot the running of $\tilde{\lambda}$ as a function of the renormalization scale for illustrative values of  
Higgs mass ($m_h$), top mass ($m_t$) and strong coupling constant 
($\alpha_s=g_3^2/4\pi$). The running of $\lambda$ is also given in the same figure for comparison.
The allowed range of values of $m_t$ ($173.2 \pm 0.9$ GeV) is taken 
from \cite{topmass} and that of $\alpha_s$ ($0.1184 \pm 0.0007$) is taken from \cite{alphas}. The Higgs
mass has been varied between $125 - 126$ GeV \cite{higgsmass}. 

We have included the corrections to incorporate the mismatch 
between the top pole mass  and $\overline{MS}$ renormalized coupling. 
This is given  
as \cite{lindner}, 
\begin{equation} 
y_t (m_t) =\frac{ \sqrt{2}m_t}{v} ( 1 + \delta_t (m_t) )  
\end{equation}
$\delta_t (m_t)$ denotes the matching correction at top pole mass. 
We include the QCD corrections 
up to three loops \cite{qcd} 
while electroweak corrections are taken up to $1$ loop \cite{matching,Schrempp:1996fb}. 
We have also included $\mathcal{O}(\alpha\alpha_s)$ correction to the matching
of top Yukawa and top pole mass \cite{Jegerlehner:2003py,bezrukov}. This correction is comparable to QCD
correction.

Suitable matching corrections for 
$\overline{MS}$ renormalized 
$\lambda$ and the Higgs 
mass at $\mu = m_t$ has been taken up to two loops \cite{Sirlin:1985ux,Degrassi:2012ry}.
The threshold effect due to the top mass is included. 
The plots corroborate the fact that for lower values of Higgs mass 
in the range reported by ATLAS and CMS, 
the stability of the vacuum 
till the Planck scale is highly restrictive \cite{Degrassi:2012ry}.
However as demonstrated in the plots, $\tilde{\lambda}\left(M_{pl}\right)$ is
not too negative. In this region the potential develops a 
new minimum from which the transition probability through quantum tunneling is
not large enough and consequently the life time of the vacuum remains
higher than the age of 
the universe. This implies that the vacuum is metastable.
The tunneling probability (at zero temperature) 
is given by \cite{Isidori:2001bm,Espinosa:2007qp}
\begin{eqnarray} 
p=\underset{\mu<\Lambda}{\text{max}}~V_U \mu^4 \,\text{exp}\left(-\frac{8\pi^2}{3|\lambda(\mu)|}\right),
\end{eqnarray} 
$\Lambda$ is the cutoff scale. $V_U$ is volume of the past light-cone which is taken as $\tau^4$ where $\tau$ is the age of the universe. 
$\tau=4.35\times 10^{17}$ sec \cite{Ade:2013ktc}. The metastability condition implies $p<1$ which can be translated to a lower bound on $\lambda$ as follows
\begin{eqnarray} 
\left|\lambda\right|<\lambda_{\text{meta}}^{\text{max}}=\frac{8\pi^2}{3}\frac{1}{4\,\text{ln}\left(\tau\mu\right)}.
\label{lam_meta}
\end{eqnarray}
This is shown in Fig.~\ref{lam-vs-mu} as a slanting line dividing the metastability and the instability region.
\subsection{Vacuum Stability and Metastability in the Minimal Linear Seesaw Model} 
\begin{figure} [ht]
\begin{center}
\includegraphics[width=8.0cm,height=5.5cm]{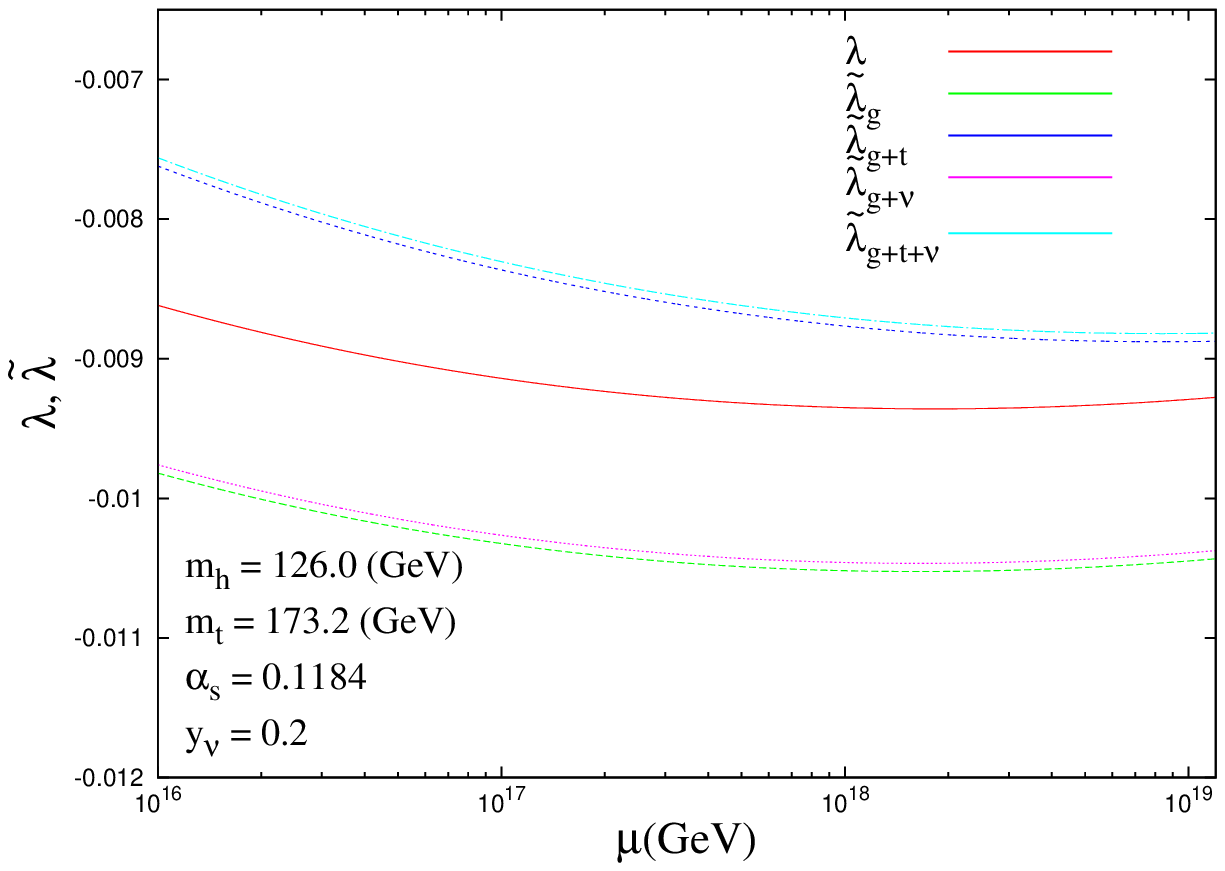}
\includegraphics[width=8.0cm,height=5.5cm]{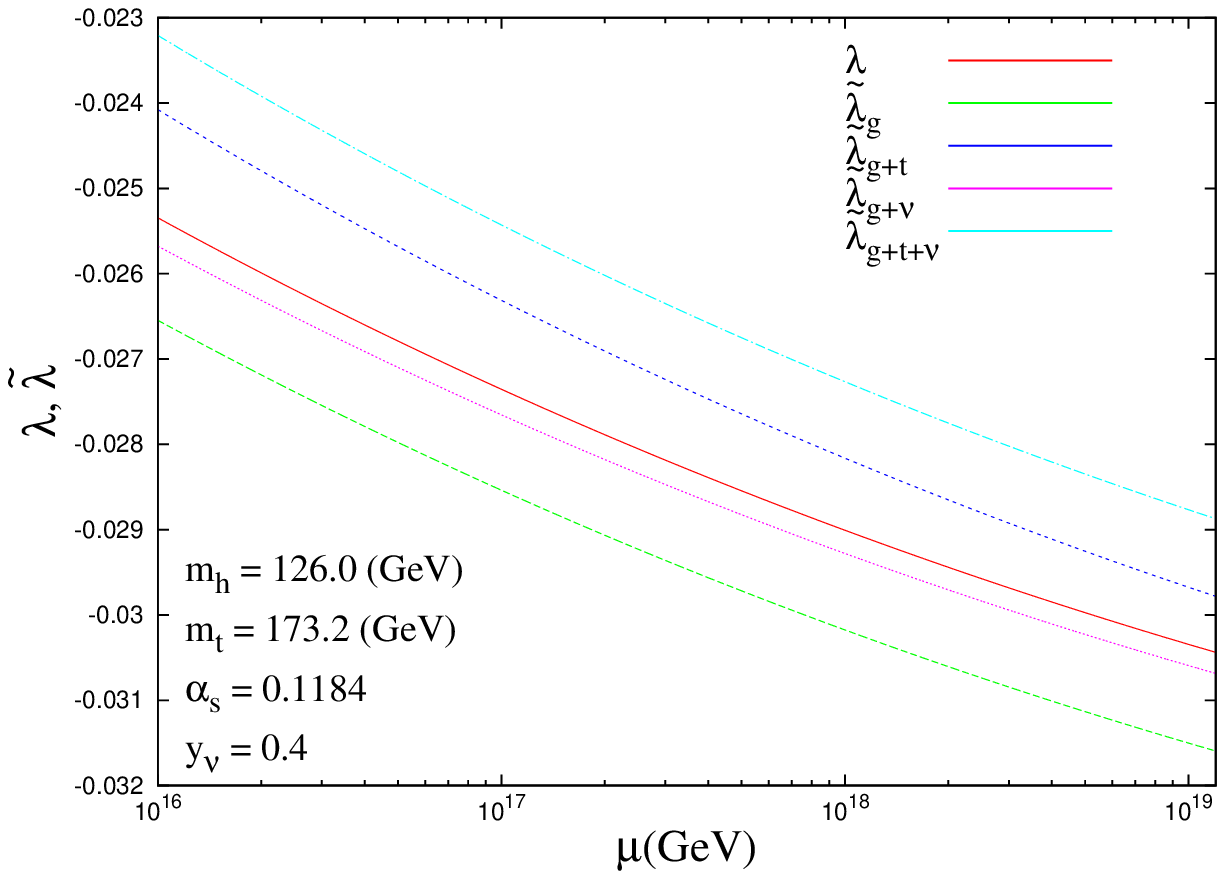} 
\caption{ Effect of the contribution from the gauge part, top and 
neutrinos to the effective coupling $\tilde{\lambda}$}
\label{effpot-nu}
\end{center}
\end{figure}

In presence of extra singlets the one loop effective potential of the 
Higgs field gets an extra contribution from the neutrinos 
\cite{Casas:1999cd}.
Generalizing the expression in \cite{Casas:1999cd}
for multi-generation case the additional part of the effective potential $V_\nu$ can be expressed as, 
\beqa
{V_\nu }\left(\phi_{cl}\right) &=& 
-\,\frac{\left(\frac{1}{2}\,\phi_{cl}^2\,
\left(Y^{\dag} Y \right)_{ii}\right)^2}{32\,\pi^2}\left[{\rm log}
\left(\frac{\frac{1}{2}\,\phi_{cl}^2\,\left(Y^{\dag} Y\right)_{ii}}
{\mu^2}\right) - \frac{3}{2}\right] \nonumber \\
& & -\,\frac{\left(\frac{1}{2}\,\phi_{cl}^2\,
\left(Y\,Y^{\dag}\right)_{jj}\right)^2}{32\,\pi^2}\left[{\rm log}
\left(\frac{\frac{1}{2}\,\phi_{cl}^2\,\left(YY^{\dag}\right)_{jj}}
{\mu^2}\right) - \frac{3}{2}\right]
\eeqa
where $\phi_{cl}$ denotes the classical value of the Higgs field.  
This is obtained in the limit $\phi_{cl} >>$ the mass of the singlet field. 
$Y$ denotes the $n \times 3$ Yukawa coupling matrix.  
Note that the above equation is in the diagonal basis for $Y^\dag Y$ and $Y Y^\dag$. 
The first contribution in the above expression comes from the light neutrinos
($i=1,2,3$) 
while the second one comes from the heavy neutrinos ($j= 1,...,n$) for 
$n$ heavy neutrinos.   
This gives rise to an additional contribution (at 1-loop) towards  
the effective self coupling $\tilde{\lambda}$  
\beqa
\tilde{\lambda}_\nu &=& -\, \frac{1}{32\,\pi^2} \left[ 
\left(\left(Y^{\dag} Y\right)_{jj}\right)^2\left(
{\rm ln}\,\frac{\left(Y^{\dag} Y\right)_{jj}}{2} - 1\right)
+ \left(\left(Y \, Y^{\dag}\right)_{jj}\right)^2
\left(
{\rm ln}\,\frac{\left(Y \, Y^{\dag}\right)_{jj}}{2} - 1\right)
\right]
\eeqa
This is to be added to the right hand side of Eq. \ref{lambdatilda}. 
In our case, $Y = Y_\nu^{\prime} = \sqrt{2} m_D^\prime/v$, $n=2$,
$(Y^{\dag} Y)^{dia} \approx Dia(0,0,y_\nu^2)$ and 
$(YY^{\dag})^{dia} \approx Dia(0, y_\nu^2)$   
in the limit of $y_s << y_\nu$.

The presence of the singlet fields also modify the 
SM Renormalization Group Equations (RGEs) for the Yukawa couplings and the Higgs 
self coupling, for  energies 
higher than the mass of the singlets.  
Including the corrections due to the neutrino Yukawa couplings up to 
one loop, the modified $\beta$ function governing the running of 
$\lambda$ is given as,  
\begin{eqnarray} 
{\beta^\prime}_\lambda^{(1)} &  = & \beta_{\lambda}^{(1)} 
+ 4 {\mathrm{Tr}({Y_\nu}^{\prime \dagger} Y_{\nu}^{\prime}}) \lambda
- 2 {\mathrm {Tr}}[ ({Y_\nu}^{\prime \dagger} Y_{\nu}^{\prime})^2 ], 
\end{eqnarray} 
where $Y_{\nu}^{\prime T}=\left(Y_{\nu}^{T},~Y_{S}^{T}\right)$. The one loop $\beta$ functions corresponding to the Yukawa couplings
$Y_u$, $Y_d$ and $Y_l$ also acquire additional factors containing 
${Y_\nu}^{\prime \dagger} Y_{\nu}^{\prime}$  \cite{1-loop-rg}. 
Finally, one needs to include the RG running of the coupling $Y_\nu^{\prime}$ 
which is governed by the following equation:
\begin{eqnarray} 
16 \pi^2 \mu \frac{d Y_{\nu}^{\prime}}{d \mu} = {Y_{\nu}^{\prime}} \left[ \frac{3}{2} {Y_\nu}^{\prime \dagger} Y_{\nu}^{\prime} 
- \frac{3}{2} {Y_l}^\dagger Y_l + T - \frac{9}{20} g_1^2-\frac{9}{4} g_2^2 \right]. 
\end{eqnarray} 
In this case  the quantity $T$ is defined as, 
\begin{eqnarray} 
T & = & {\mathrm {Tr}} \left[3 {Y_u}^\dagger Y_u + 3 {Y_d}^\dagger Y_d + {Y_l}^\dagger Y_l +  {Y_\nu}^{\prime \dagger} Y_{\nu}^{\prime}\right].   
\end{eqnarray} 
RG equation for neutrino Yukawa coupling is taken up to one loop \cite{1-loop-rg}.

The $Y_{\nu}^{\prime}$ dependence of the beta function of $\lambda$ is in terms of
Tr$\left[{Y_\nu}^{\prime \dagger} Y_{\nu}^{\prime}\right]$ 
and Tr$\left[({Y_\nu}^{\prime \dagger} Y_{\nu}^{\prime})^2\right]$
only. From the parameterization
of $Y_{\nu}$ and $Y_S$, we find
\begin{eqnarray}
\text{Tr}\left[{Y_\nu}^{\prime \dagger} Y_{\nu}^{\prime}\right]&=&y_{\nu}^2 +y_s^2
\simeq y_{\nu}^2, \nonumber \\
\text{Tr}\left[{Y_\nu}^{\prime \dagger} Y_{\nu}^{\prime}{Y_\nu}^{\prime \dagger} Y_{\nu}^{\prime}\right]&=& y_{\nu}^4 + 
2 y_{\nu}^2y_s^2\rho^2 + y_s^4 
\simeq y_{\nu}^4,
\label{tr_ynu}
\end{eqnarray}
\begin{figure} [ht]
\begin{center}
\includegraphics[width=8.0cm,height=5.5cm]{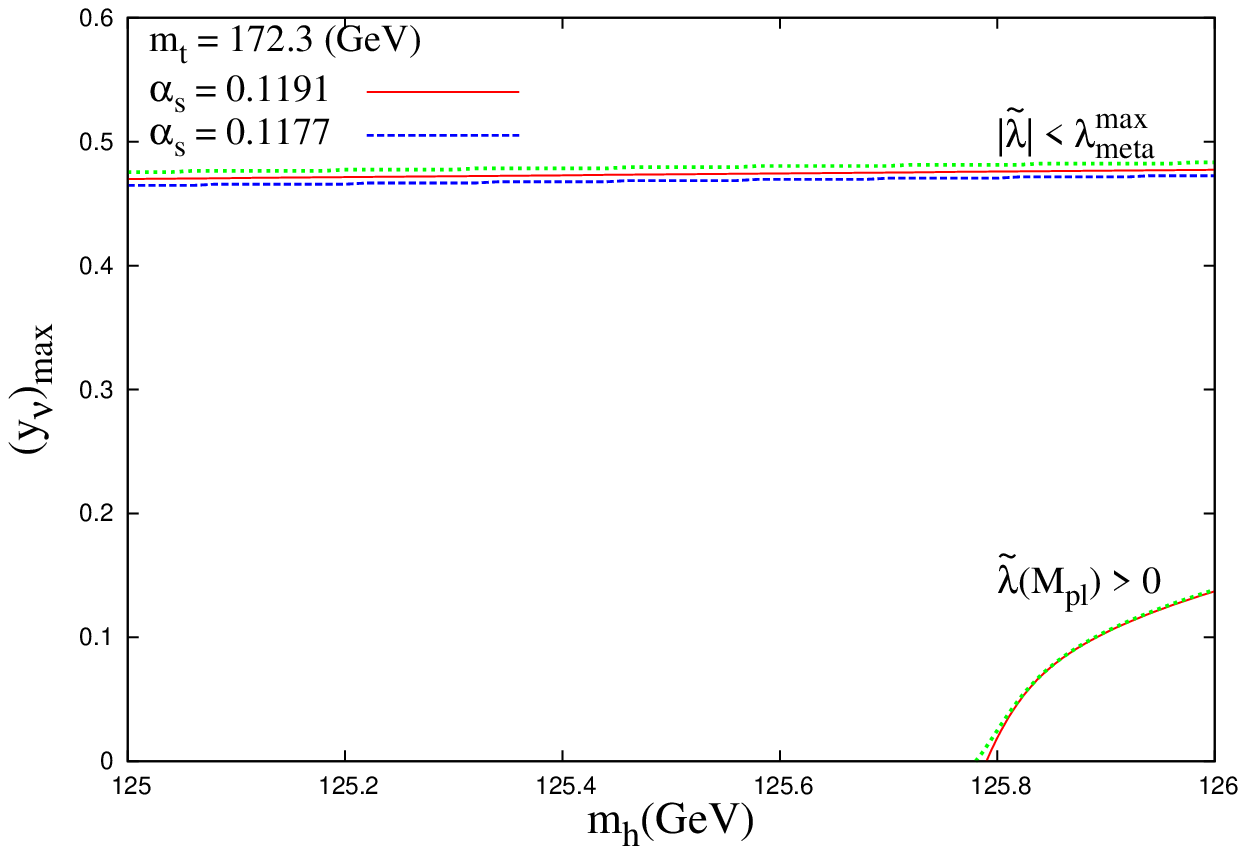}
\includegraphics[width=8.0cm,height=5.5cm]{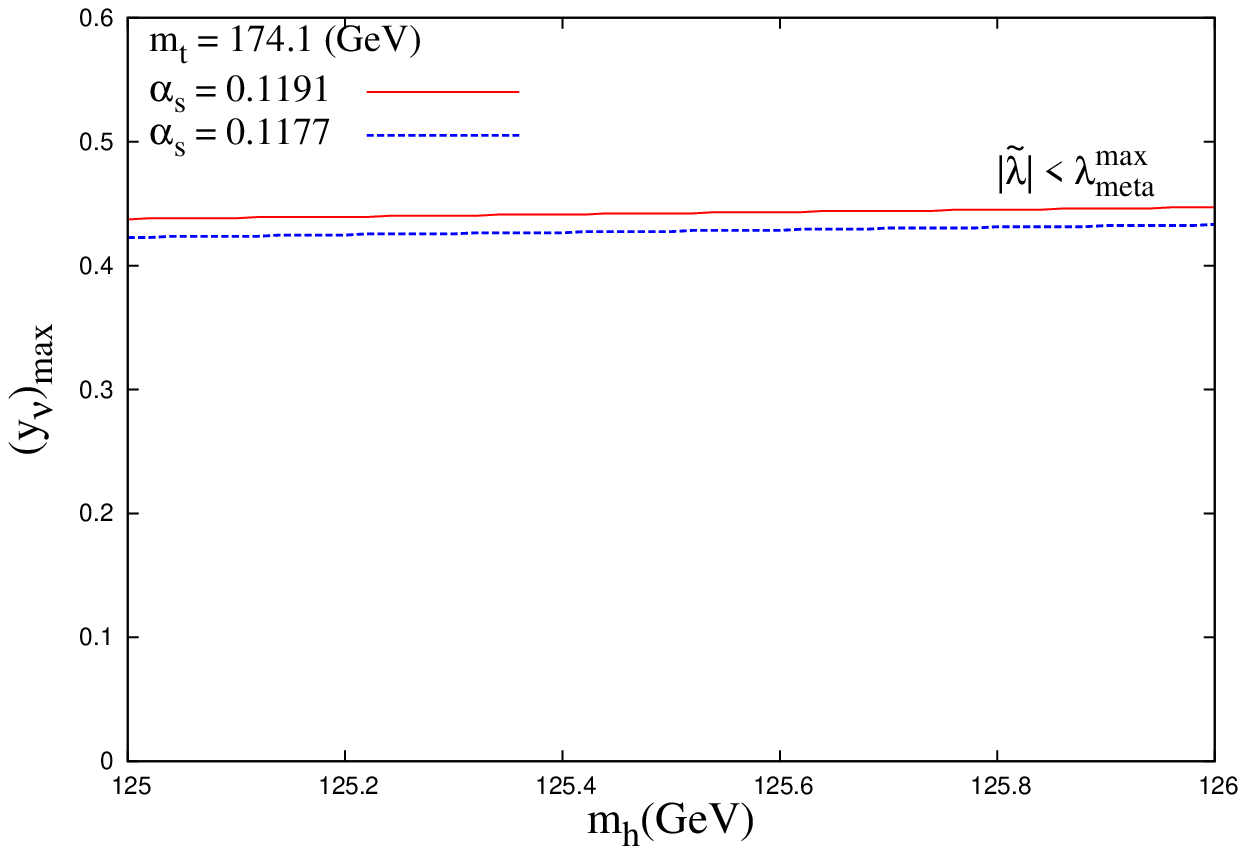} 
\caption{ The allowed region of $y_{\nu}$ with varying Higgs mass from the consideration of meta stability 
$\left(\left|\tilde{\lambda}\right|<\lambda_{\text{meta}}^{\text{max}}\right)$ and absolute stability 
$\left(\tilde{\lambda}(M_{pl})\ge 0\right)$.
The region below the curves is allowed. 
}
\label{alow-rgn-ynu-higgs}
\end{center}
\end{figure}
\begin{figure} [ht]
\begin{center}
\includegraphics[width=8.0cm,height=5.5cm]{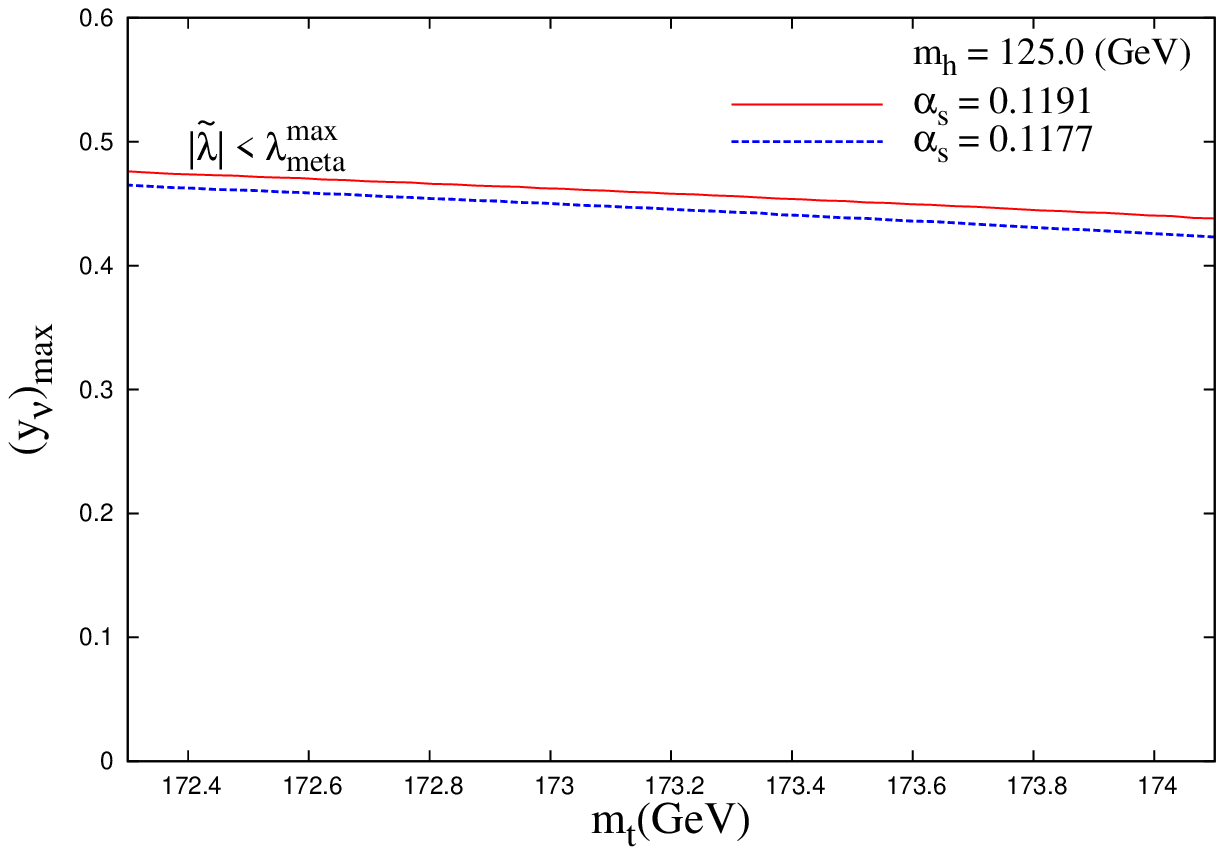}
\includegraphics[width=8.0cm,height=5.5cm]{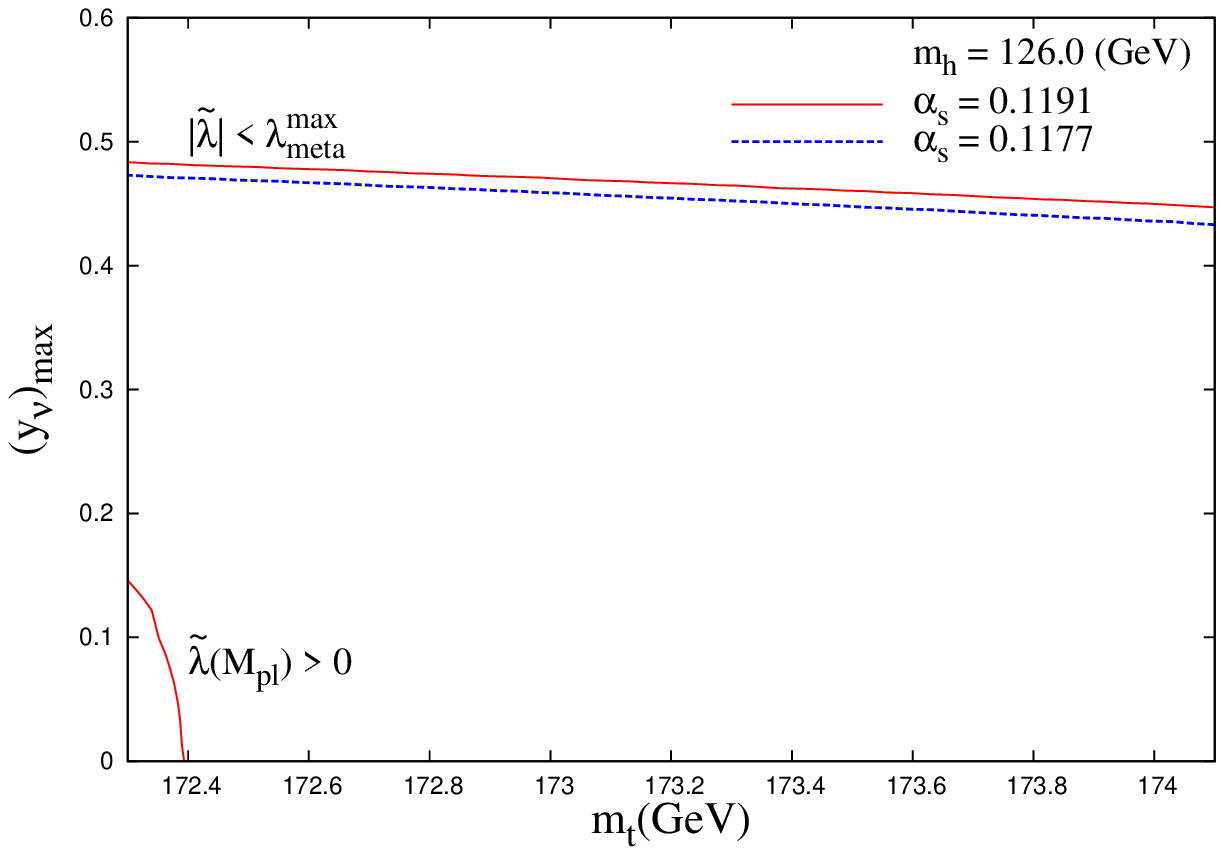} 
\caption{ Same as in Fig. \ref{alow-rgn-ynu-higgs} with varying top mass.}
\label{alow-rgn-ynu-top}
\end{center}
\end{figure}
since $y_s << y_\nu$. The exact equalities in the above expressions are also valid even without the parameterization 
of Eq.~(\ref{pmtz-nh},\ref{pmtz-ih}). However $\rho$ remains undetermined. Also, 
in this case the smallness of $y_s$ makes the trace terms to be dependent only on $y_{\nu}$
and hence $\beta_\lambda$ depends on  only one unknown parameter ({\it i.e.}$\,y_\nu$).
Also we can see from Eq.~(\ref{tr_ynu}) that the trace terms do not depend on the neutrino oscillation parameters. 
In addition, under the approximation of $y_s << y_\nu$, there is no dependence on mass hierarchy as well 
(only $\rho$ depends on hierarchy).

In Fig. \ref{lam-vs-mu} we show the effect of inclusion of this term on the running of ${\lambda}$. 
As expected, $\tilde{\lambda}$ becomes more negative near Planck scale  
in presence of the seesaw term. The figures are obtained including the heavy 
neutrino contribution to the effective potential.

In Fig. \ref{effpot-nu} we explicitly display the effect of inclusion of the neutrino 
contribution in the effective potential. The left panel is for $y_\nu = 0.2$ 
and the right panel is for $y_\nu=0.4$. 
The solid-red  curve shows the self-coupling $\lambda$. Comparing the 
two plots for this in the two panels we see that the higher value of $y_\nu$ 
makes $\lambda$ more negative in the right panel. 
The dashed-green curve shows the evolution of the effective self-coupling 
due to the contribution of the gauge fields while the small-dashed-blue curve 
shows the effect of inclusion of the contribution from the top quarks. 
We see that since in the expression of effective coupling the gauge contribution
and top contribution come with opposite signs, these terms drive $\tilde{\lambda}$ 
in opposite directions. While the gauge-contribution makes the 
effective $\lambda$ more negative, the top contribution makes it more positive.
The dot-dashed-cyan lines show the contributions from the neutrino. 
This effect goes in the same direction as the top contribution. 
The effect is found to be very small for $y_\nu=0.2$. 
However for larger values of $y_\nu$ the effect can be non-negligible 
as can be seen from the right panel for $y_\nu=0.4$.

In Fig. \ref{alow-rgn-ynu-higgs} we give the plot of the allowed region
of $y_{\nu}$ as a function of the Higgs mass for fixed values of top mass  
and the strong coupling constant. 
The slanting lines are obtained by imposing the condition given in Eq.~(\ref{lam_meta}).
$y_\nu=0$ corresponds to the SM. 
The first panel is for $m_t = 172.3$ GeV. For this value of top mass and $\alpha_s = 0.1191$, 
only a tiny parameter space is allowed through the stability condition ({\it i.e.} $\tilde{\lambda}(M_{pl})\ge 0$). 
Inclusion of the neutrino contribution to the effective potential, shown by the 
dashed-green lines, 
does not have any discernible effect on the stability bound of $\tilde{\lambda}$ for the values of 
$y_\nu$ concerned. 
However metastability bound changes by $\sim\,1\%$ as is visible from the figure. The solid-red 
line is the metastability bound with $\lambda$. The dashed-green line is with $\tilde{\lambda}$ 
including the neutrino effect. All other metastability bounds including the dashed-blue one in 
this figure are plotted for $\tilde{\lambda}$ including the neutrino effect. 
Inclusion of heavy neutrino contribution in $\tilde{\lambda}$ changes the bound 
on $y_\nu \le 0.483$ as opposed to $y_\nu \le 0.477$ obtained without including the 
neutrino Yukawa term.
The $2^{\text{nd}}$ panel corresponds to a higher value of $m_t = 174.1$ GeV. 
In this case we obtain the upper bound $y_\nu \ls 0.45$ depending on the 
value of $\alpha_s$. This bound is obtained for a Higgs 
mass of 126 GeV. For lower values of Higgs mass, the allowed $y_\nu$ values 
are correspondingly lower. 

\begin{figure} [ht]
\begin{center}
\includegraphics[width=8.0cm,height=5.5cm]{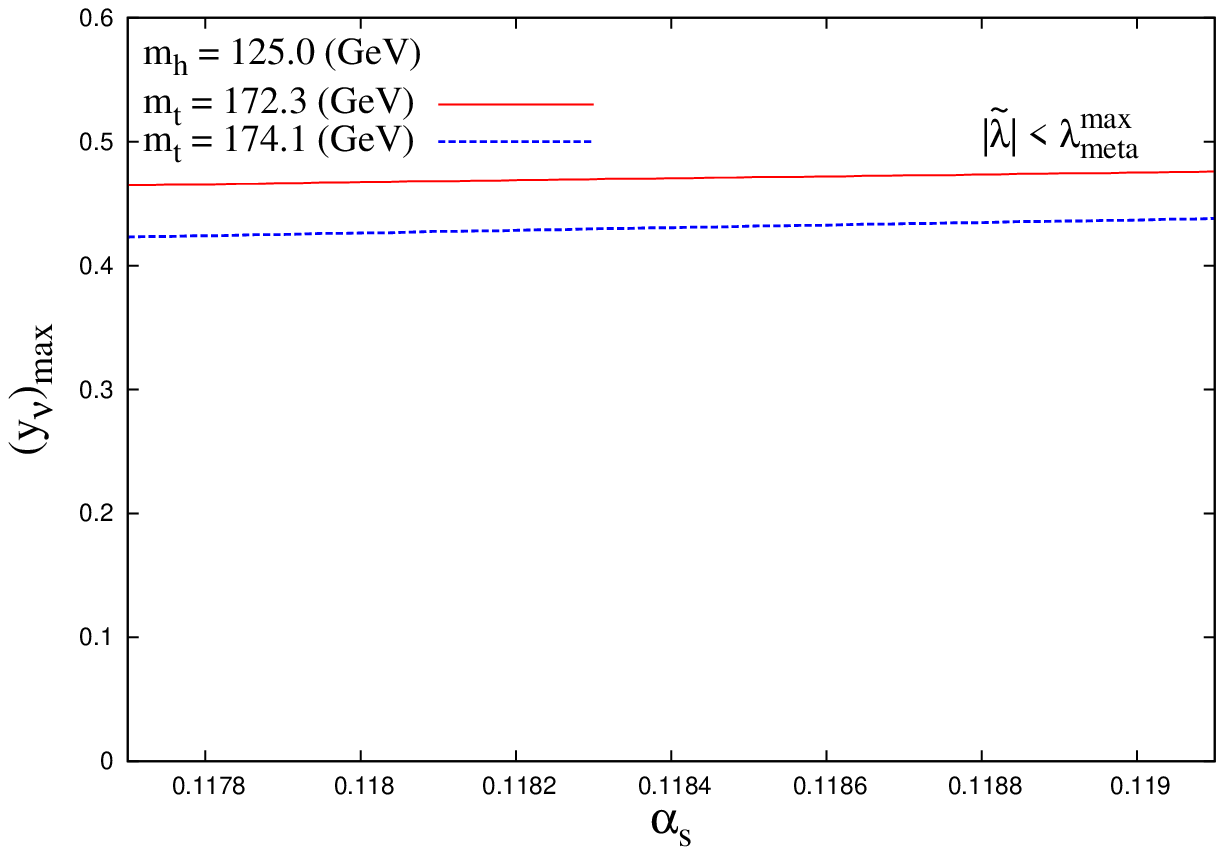}
\includegraphics[width=8.0cm,height=5.5cm]{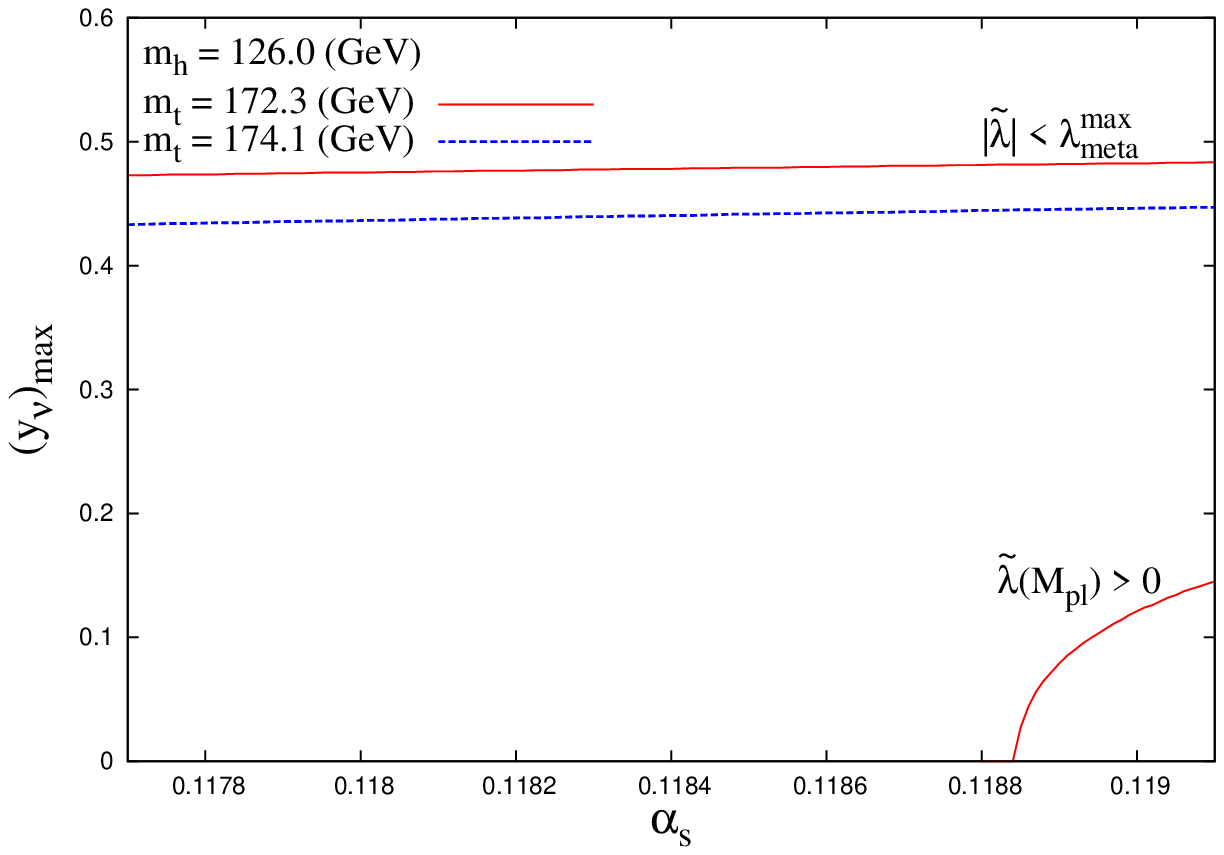}
\caption{Same as in Fig. \ref{alow-rgn-ynu-higgs} with varying strong coupling constant.}
\label{alow-rgn-ynu-alphas}
\end{center}
\end{figure}

Fig. \ref{alow-rgn-ynu-top} 
shows the variation of $y_\nu$ with $m_t$ for fixed values of Higgs mass. 
Fig. \ref{alow-rgn-ynu-alphas}  show the variation of $y_\nu$  with 
$\alpha_s$. In general, slightly larger allowed regions are obtained for
higher values of Higgs mass, lower values of top mass and higher values of 
$\alpha_s$.   
From the above figures we can have an overall  upper bound on the value of
$y_{\nu}$
as
\be
y_{\nu} \ls 0.48.
\ee
The metastability condition is imposed on the value of $\tilde{\lambda}$ by 
replacing $\lambda$ with $\tilde{\lambda}$ in Eq.~(\ref{lam_meta}).
\begin{figure} [ht]
\begin{center}
\includegraphics[width=8.0cm,height=5.5cm]{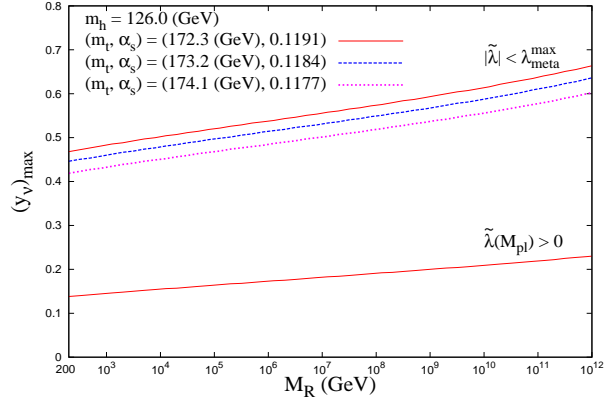}
\caption{The left-panel shows the
upper bound on $y_{\nu}$ as a function of the
right-handed neutrino mass from the consideration of metastability $\left(\left|\tilde{\lambda}\right|<\lambda_{\text{meta}}^{\text{max}}\right)$ 
and absolute stability $\left(\tilde{\lambda}(M_{pl})\ge 0\right)$. For absolute stability there is only one line corresponding 
to $(m_t,\alpha_s)=(172.3 \,\text{GeV},0.1191)$ as for other set of parameters no allowed region is available.
}
\label{ynuvsmr}
\end{center}
\end{figure}

The above plots are obtained by keeping $M_R$ fixed at 1 TeV. 
In Fig. \ref{ynuvsmr} we show how the upper bound on $y_\nu$ depends on 
the scale of $M_R$.

We see that variation of $M_R$ within a few TeV (which is our range of 
current interest) would not change the 
bound on $y_\nu$ drastically.  In fact this trend continues 
even if $M_R$ is increased to higher values. At $M_R = 10^{12}$ GeV, the upper bound
on $y_{\nu}$ obtained is, $y_{\nu} \ls 0.67$. 
Note that the limits on neutrino mass squared differences from oscillation
experiments constraint the product of $y_\nu$ and $y_s$ as 
$y_\nu y_s \approx M_R \sqrt{\Delta m^2_{atm}}/v^2$  
for 
both hierarchies   
in the limit 
${\Delta m_{\odot}^2}/{\Delta m^2_{atm}} << 1$. 
Thus, for 
higher values of $M_R$, $y_s$ needs to be increased to 
keep the neutrino mass in the desired range. 
Beyond $10^{12}$ GeV the contribution 
from the $y_s$ term starts getting significant  
and hence this needs to be included in the RG evolution. 

\section{Non-unitarity of Neutrino Mixing Matrix and Lepton Flavor Violation} 
The mechanism of neutrino oscillation has already indicated
that there is flavor violation in the lepton sector. 
The question arises if there can be flavor violating decays 
in the charged lepton sector. In typical Seesaw models this rate is very small 
because of the smallness of the light-heavy mixing. 
However in TeV scale seesaw models, since lepton number violation (LNV)
is separated from the scale of lepton flavor violation (LFV), this may not be 
the case \cite{tevlfv}. 
It is well known that LNV is due to the dimension 5 operator whereas the
LFV can be related to the dimension 6  operator. 
In general, the flavor structure of the coupling strengths of these operators 
are not correlated. However, in MLSM  such a relation can be established  from the 
hypothesis of minimal flavor violation \cite{MFV}.    
In this model, the mass and gauge eigenstates are related as, 
\be
\begin{pmatrix}
\nu_L & \nu^c  
\end{pmatrix} 
= U 
\begin{pmatrix}
{\nu_L}^\prime & {\nu^c}^\prime 
\end{pmatrix} 
\ee 
where,  
\be
\nu^c = 
\begin{pmatrix}
N_R^c & S^c
\end{pmatrix}. 
\ee
The leptonic part of the charged current interaction in the gauge basis
is given as ,
\begin{eqnarray}
\mathcal{L}_{\rm CC} &=& \frac{g}{\sqrt{2}}\, \sum_{\alpha=e, \mu, \tau}
\overline{\ell}_{\alpha \,L}\, \gamma_\mu {\nu}_{\alpha \,L}\, W^{\mu} + \text{h.c.} 
\end{eqnarray}
This can be expressed in the mass basis as,
\begin{eqnarray}
\mathcal{L}_{\rm CC} &=& \frac{g}{\sqrt{2}}\, \sum_{\alpha=e, \mu, \tau}\, 
\sum_{i=1}^{3} \sum_{j=1}^{2}
\left[ \overline{\ell^\prime}_{\alpha \,L}\, \gamma_\mu\,  U_l^\dagger
                    \{(U_L)_{\alpha i} \nu_{L i}^\prime+(V)_{\alpha j} N_{R j}^c\} W^{\mu}
\right]
 + \text{h.c.}  
\end{eqnarray}
The PMNS matrix  is defined as
\begin{eqnarray}
\label{upmns}
U_{\text{PMNS}}=U_{l}^{\dag}\left(1-\frac{1}{2}\epsilon\right)U
\end{eqnarray}
where $U_{l}$ is the unitary matrix which takes the left-handed charged lepton
fields to their mass basis and other quantities are defined earlier. As we are working 
in a basis, where charged lepton mass matrix is diagonal, $U_l$ is being taken as unity.
We note that $U_{PMNS}$ is non-unitary and the correction to unitarity is 
proportional to $\epsilon/2$. 
\begin{table}[t]
\caption{Various experimental constraints from charged lepton flavor violating decays \cite{pdg-lfv}.
\label{table-lfv}}
\begin{ruledtabular}
\begin{tabular}{cc}
Branching Ratios & Experimental constraints \\
\hline
Br($\mu \rightarrow e \gamma$) & $< 5.7 \times 10^{-13}$ \\
Br($\tau \rightarrow e \gamma$) & $< 3.3 \times 10^{-8}$ \\
Br($\tau \rightarrow \mu \gamma$) & $< 4.4 \times 10^{-8}$ \\
Br($\mu \rightarrow 3e$) & $< 1.0 \times 10^{-12}$ \\
Br($\tau \rightarrow 3e$) & $< 2.7 \times 10^{-8}$ \\
Br($\tau \rightarrow 3\mu$) & $< 2.1 \times 10^{-8}$ \\
Br($\tau \rightarrow e \mu \mu$) & $< 1.7 \times 10^{-8}$ \\
Br($\tau \rightarrow ee \mu$) & $< 1.5 \times 10^{-8}$ \\
\hline
\end{tabular}
\end{ruledtabular}
\end{table}

In this section we consider the branching ratios of LFV decays in MLSM. 
In view of the recent measurement of $\theta_{13}$ the branching ratios 
now can be studied in terms of the CP phases.  
In addition, from the experimental upper limits on LFV processes 
one can  obtain constraints on the parameter $y_\nu/M_R$ as a function of the CP
phases. When combined with the  upper bound on $y_\nu$ from 
vacuum (meta)stability  as a function of $M_R$, the parameter space can be further constrained. 
Table \ref{table-lfv} lists the experimental constraints coming from the charged lepton
flavor violating decays \cite{pdg-lfv}.

Nevertheless, in this section we will concentrate only on the constraints coming from Br($\mu \rightarrow e \gamma$) since this
is the most constraining as can be seen from Table \ref{table-lfv}.

Branching ratio for the process, $\mu \rightarrow e\gamma$ is given by \cite{lfv}
\begin{eqnarray}
\text{Br}\left(\mu \rightarrow e\gamma\right) = \frac{3\alpha}{8\pi}\left|V_{ei}V_{i\mu}^{\dag}f(x) \right|^2,
\label{eq:mu2egamma} 
\end{eqnarray}
where, $x = \left(\frac{M_i^2}{m_W^2}\right)$
and 
\begin{eqnarray}
f(x)=\frac{x\left(1-6x+3x^2+2x^3-6x^2\text{ln}x\right)}{2(1-x)^4}.
\end{eqnarray}
$f(x)$ is a slowly varying function of $x$ ranging from $0$ to $1$ for 
$x$ between $0$ to infinity. 
The light-heavy mixing matrix $V$ is defined in Eq.~(\ref{bdmatrix}). 
The current experimental constraint on this is \cite{mu2egamma} ~~(see, table \ref{table-lfv})
\begin{equation}
\text{Br}\left(\mu \rightarrow e\gamma\right) < 5.7 \times 10^{-13}.
\label{brmu2e} 
\end{equation} 
Using the parameterization of $Y_{\nu}$ and $Y_S$ in Eq.(\ref{pmtz-nh}) and 
(\ref{pmtz-ih}), we obtain, for normal hierarchy
\begin{eqnarray}
\text{Br}\left(\mu \rightarrow e\gamma\right) & = & 
\frac{3\alpha}{8\pi}\frac{y_{\nu}^4 v^4}{4 M_R^4}f^2(x)\left(\sqrt{r}\,s_{12}^2+2\,r^{(1/4)}\,s_{13}\,s_{12}\,s_{(\alpha+\delta)} + r^{(3/4)}\,c_{23}\,s_{12}\,s2_{12}\,s_{\alpha}/s_{23} \right)s_{23}^2  \nonumber 
\\ 
& + & \mathcal{O}\left(y_s,(\sqrt{r},s_{13})^2\right).
\end{eqnarray}
In the above expressions and in subsequent part, we have used the following notations
\begin{eqnarray}
& &s_{ij}=\text{sin}\theta_{ij}, \quad c_{ij}=\text{cos}\theta_{ij}, \quad s_{\alpha+\delta} =\text{sin}(\alpha+\delta), \quad s_{\delta} =\text{sin}\,\delta, \quad s_{\delta} =\text{sin}\,\delta \nonumber \\
& &\quad s2_{ij} = \text{sin}2\theta_{ij}, \quad c2_{ij} = \text{cos}2\theta_{ij}, \quad c_{2\alpha} =\text{cos}2\alpha, \quad s_{2\alpha} =\text{sin}2\alpha \quad \text{etc}.
\end{eqnarray}
\begin{figure} [ht]
\begin{center}
\includegraphics[width=8.0cm,height=5.5cm]{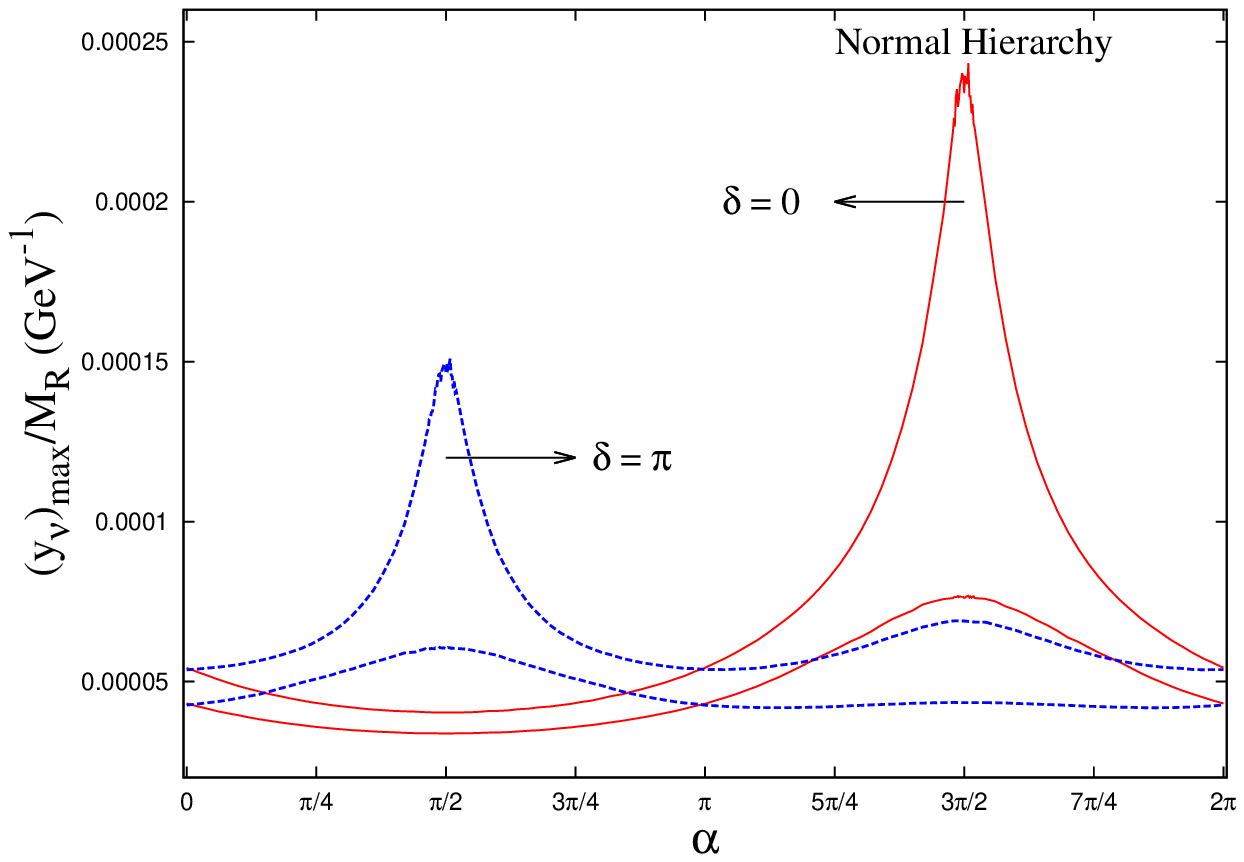}
\includegraphics[width=8.0cm,height=5.5cm]{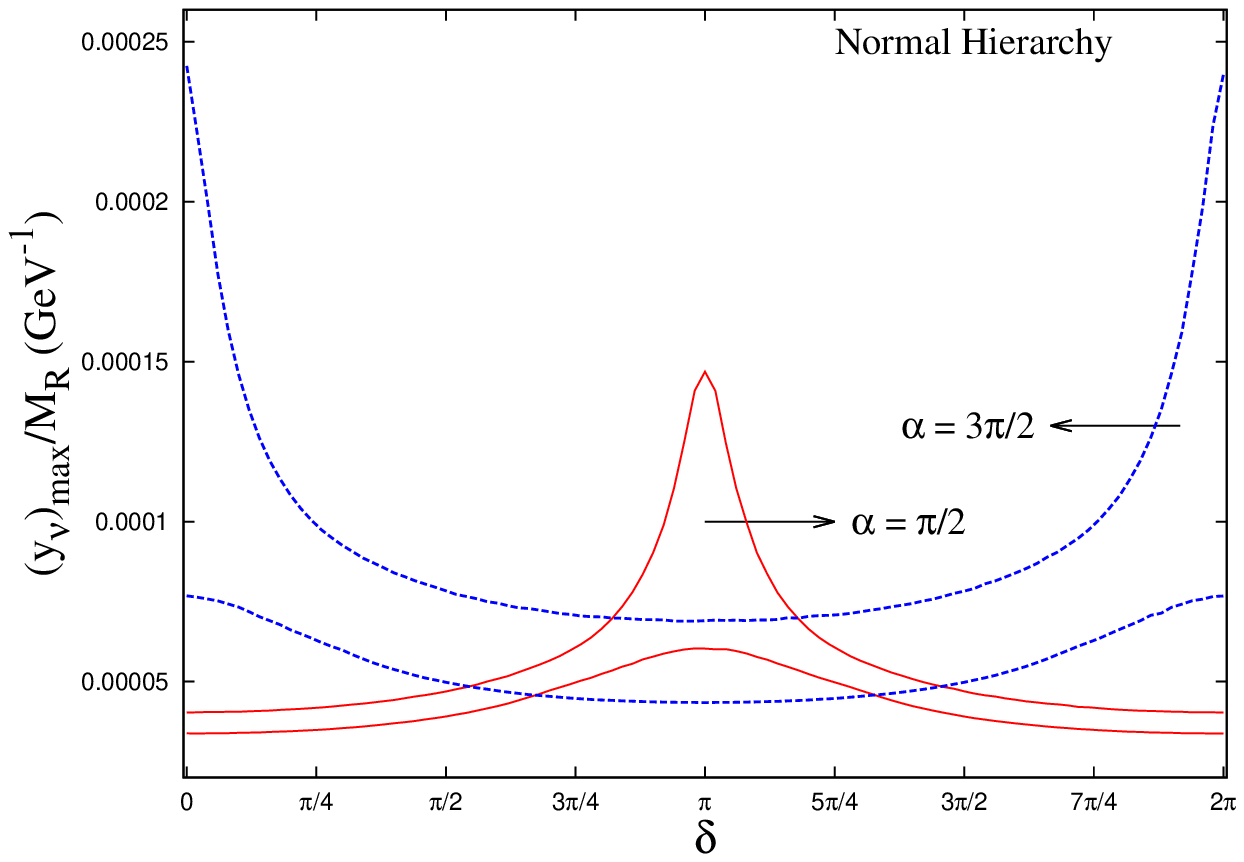} \\
\includegraphics[width=8.0cm,height=5.5cm]{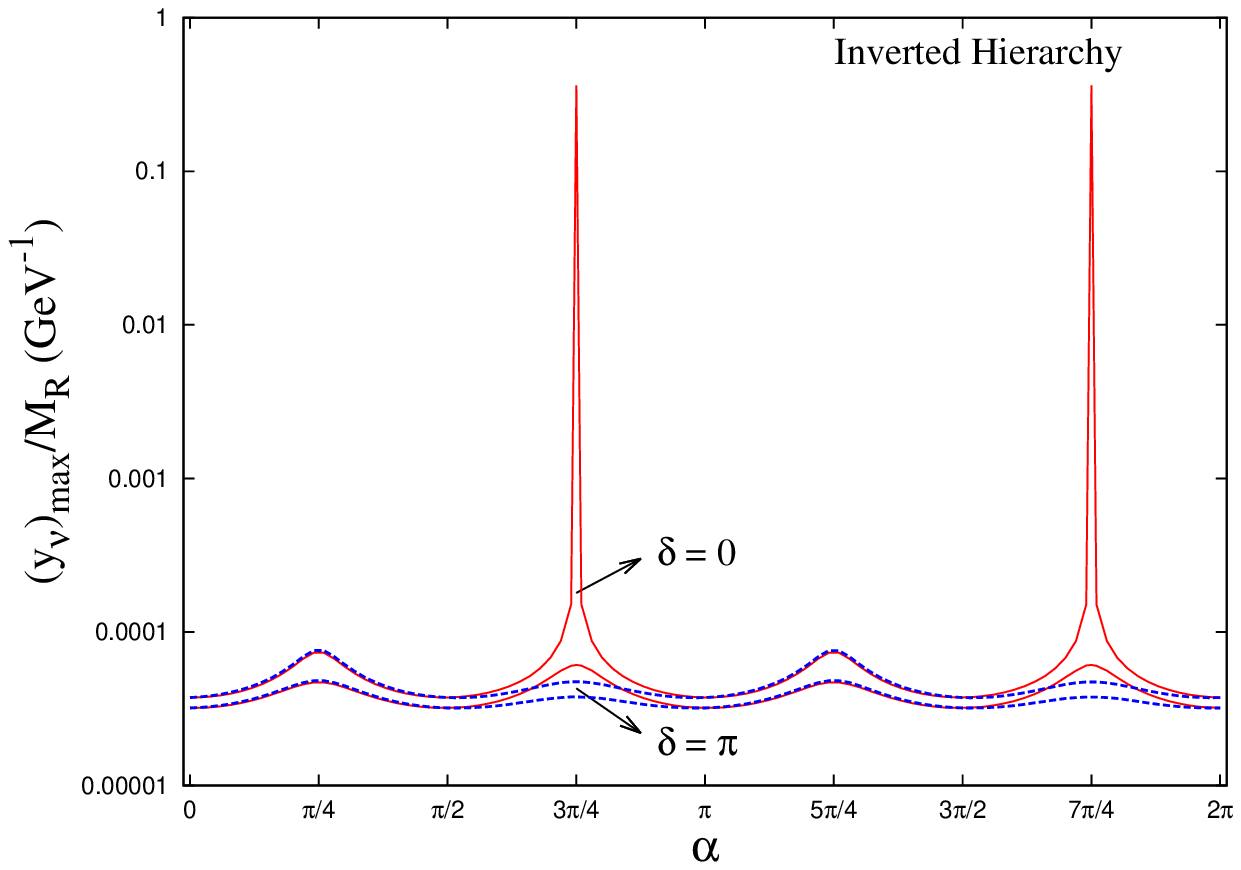}
\includegraphics[width=8.0cm,height=5.5cm]{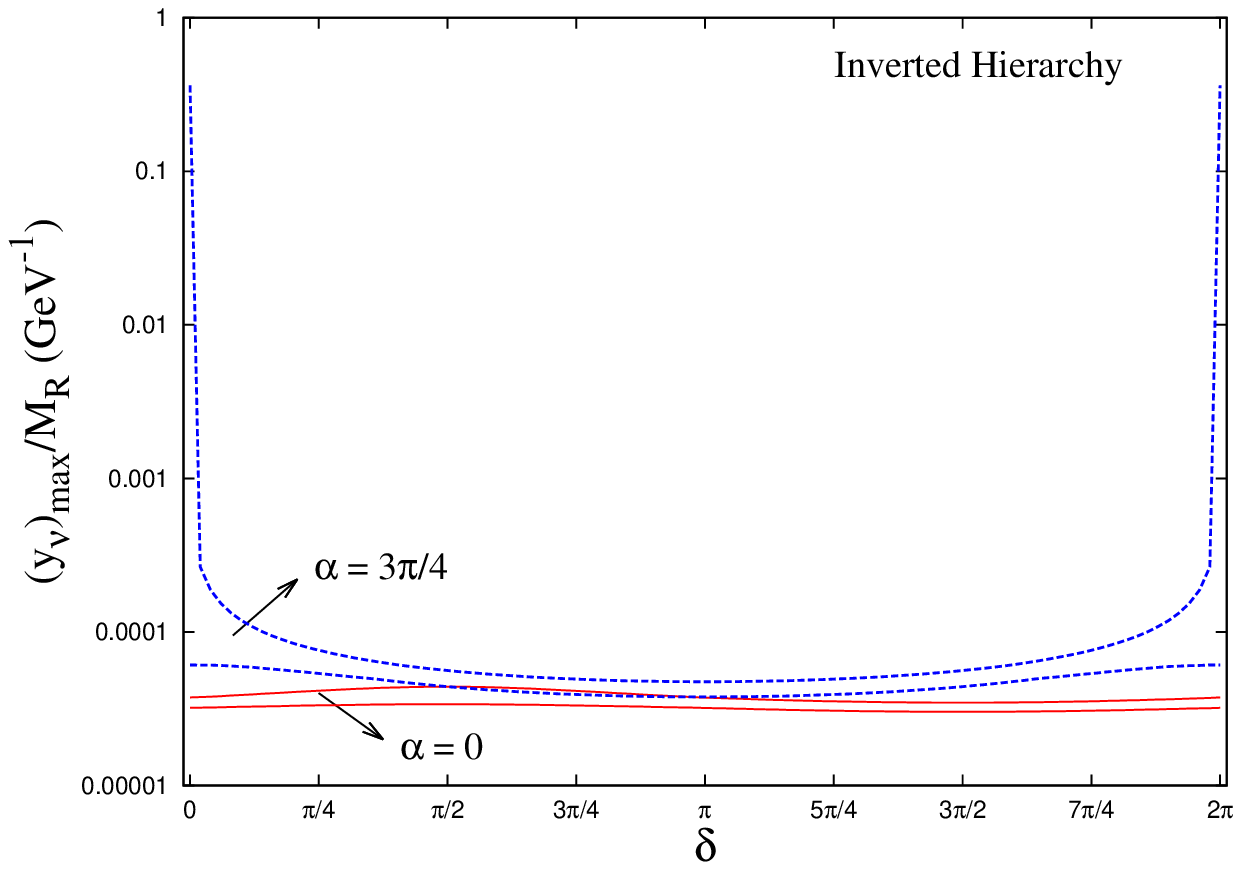}
\caption{The upper panels show the  allowed regions of
$y_\nu/M_R$ vs the CP phases $\alpha$  and $\delta$
for NH while the lower panels are for IH.
The area below each curve is consistent with the
experimental upper bound on the rate of $\mu \rightarrow e \gamma$.
The two lines of the same type (color) correspond to the maximum and minimum
value of the upper bound obtained by varying the oscillation parameters over
their $3\sigma$ range.
}
\label{ynumr-vs-phases}
\end{center}
\end{figure}

The above equation in conjunction to the upper bound on the $Br(\mu \rightarrow e \gamma)$ can be used to put an upper bound on 
$y_\nu/M_R$ as, 
\begin{equation} 
y_\nu/M_R < \left[\frac{5.7 \times 10^{-11}}{3 \alpha v^4 f^2(x)G^{NH}(r,\theta_{ij},\alpha, \delta)}\right]^{1/4}.
\label{eq:ymubymr_nh}
\end{equation}  
\begin{figure} [ht]
\begin{center}
\includegraphics[width=8.0cm,height=5.5cm]{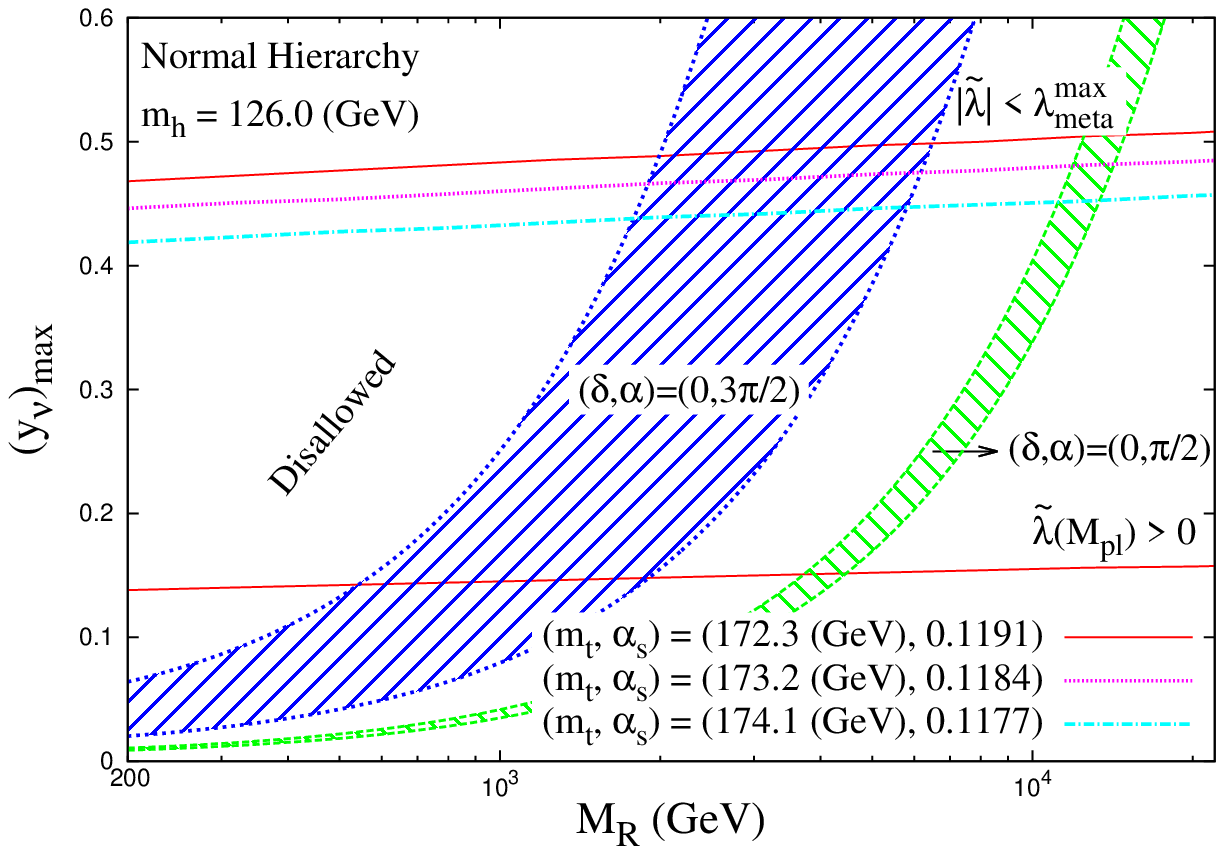}
\includegraphics[width=8.0cm,height=5.5cm]{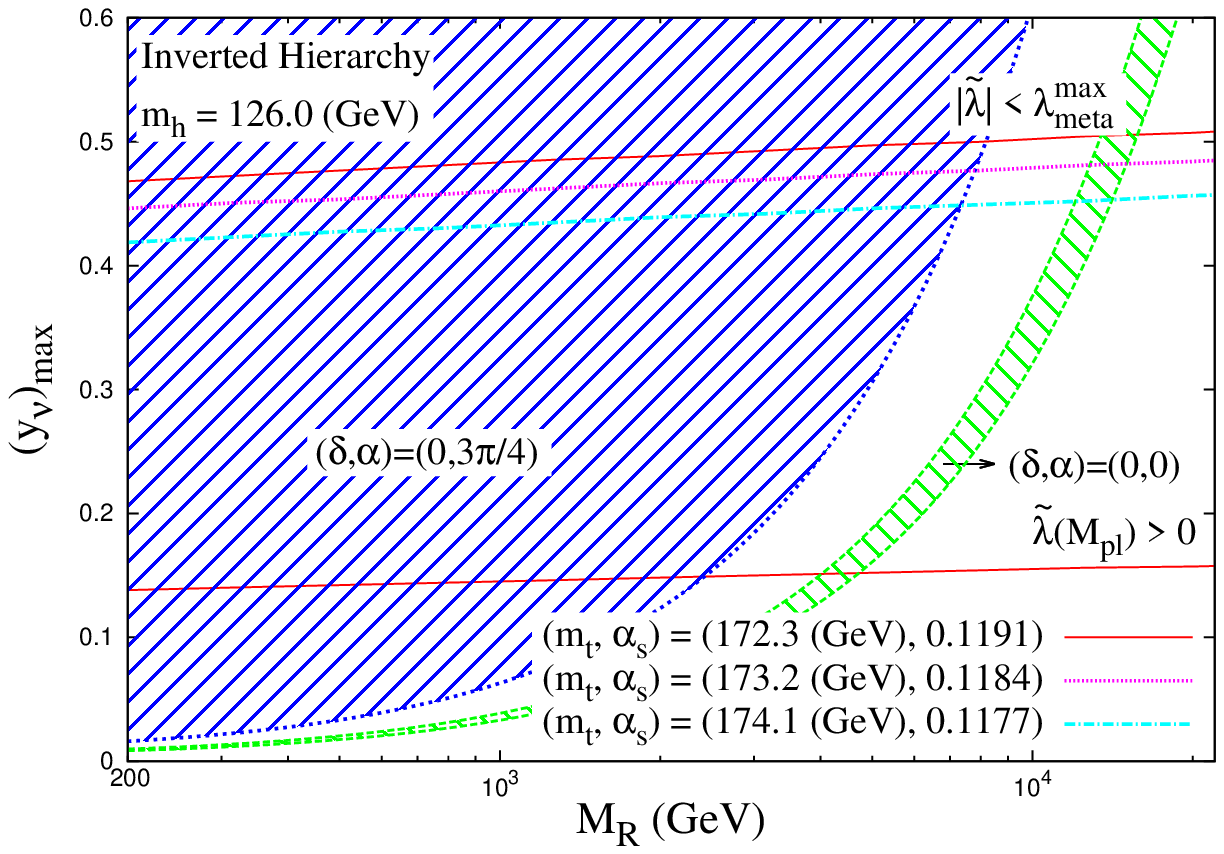}
\caption{The allowed regions of
$y_\nu^{max}$ as a function of $M_R$ from the combined constraints of $Br(\mu \rightarrow e \gamma)$ and vacuum (meta)stability. The area to the right of the curved lines are allowed from experimental bound on $Br(\mu \rightarrow e \gamma)$ while the area below the slanting lines are allowed from the constraint on vacuum (meta)stability. 
}
\label{vac-stab+lfv}
\end{center}
\end{figure}

Here, 
the factor $G^{NH}(r,\theta_{ij},\alpha, \delta)$
contains the oscillation parameters. 
The upper bound on $y_\nu/M_R$ varies in a range 
depending on the values of the CP phases.  
The minimum value of the upper bound  
occurs at $\alpha + \delta = \frac{\pi}{2}$ 
while the maximum occurs at $\alpha + \delta = \frac{3\,\pi}{2}$. 
This is reflected in the top panels in 
Fig. \ref{ynumr-vs-phases} where we display the allowed values of
$y_\nu/M_R$ as a function of the CP phases\footnote{In this plot we have taken $f(x)$ to be unity. For $M_R = 200$ GeV, there will be a multiplicative factor $\sim 1.3$. As $M_R$ increases, this factor tends to become unity. In Fig. \ref{vac-stab+lfv}, we have included the exact value of $f(x)$ at each $M_R$.}. 
The left most panel displays the variation of the 
upper bound in $y_\nu/M_R$ as a function of the Majorana phase
$\alpha$. The solid(red) line corresponds to 
the Dirac phase $\delta = 0$ while the dashed (blue) line is 
for $\delta = \pi$.  The other oscillation parameters 
are marginalized over the 3$\sigma$ range in Table 1 to give
the maximum and minimum value  of the upper bound on 
$y_\nu/M_R$. From the figure it can be inferred  that the maximum value of the 
upper bound on $y_\nu/M_R$ is 
\be
y_\nu/M_R \ls 0.00024 (\mathrm{GeV^{-1}}),
\ee
which occurs for $\delta=0$ and $\alpha=\frac{3\,\pi}{2}$ for NH.  

For IH, the branching ratio can be expressed as,  
\begin{eqnarray}
\text{Br}\left(\mu \rightarrow e\gamma\right) &=& 
\frac{3\alpha}{8\pi}\frac{y_{\nu}^4 v^4}{4 M_R^4}
f^2 \left(\frac{M_R^2}{m_W^2}\right)
\frac{1}{4}c_{23}\left(1-s_{2\alpha}s2_{12}\right)\left[c_{23}\left(1+s_{2\alpha}s2_{12}\right)\left(1-s_{13}\right)+2\left(s_{2\alpha}c_{\delta}c2_{12}-c_{2\alpha}s_{\delta}\right)s_{13}s_{23}\right]  
\nonumber \\
& & + \mathcal{O}\left(y_s,(\sqrt{r},s_{13})^2\right).
\end{eqnarray}
From this one can again get an upper bound on 
$y_\nu/M_R$. 
This is displayed in the lower panels in Fig. \ref{ynumr-vs-phases}. 
As in the case of NH, the value of the upper bound depends on the CP phases. 
The maximum allowed value in this case is 
\be 
y_\nu/M_R \ls 0.36 (\mathrm{GeV^{-1}}),
\ee
which 
occurs for $\delta =0, \alpha=\frac{3\,\pi}{4}$ as can be seen from the figure. 
The lines of same line type (color) 
corresponds to the upper bounds including the uncertainties in the masses and mixing 
parameters.  

The maximum  value of  $y_\nu/M_R$ as obtained above can be used to  retrieve  
the maximum value of $y_\nu$ for each $M_R$. This is shown in Fig. \ref{vac-stab+lfv}. 
Note that while extracting the bound on $y_\nu$ for a particular $M_R$ from the figure
one has to be careful to ensure that the perturbativity bound on $y_{\nu}$ ($\ls 1$) is 
not violated.
We also superimpose  the bounds obtained from consideration of 
vacuum (meta)stability in this figure.   
The area to the left of the shaded bands is disallowed from 
the constraint on the branching ratio 
$\mu \rightarrow e \gamma$. These bands are obtained for 
fixed values of the CP phases $(\delta,\alpha)$. The
band for each combination of 
CP phase is obtained by varying the 
oscillation parameters in their current 3$\sigma$ range. 
The area below the slanting lines 
are allowed from the constraint on vacuum  (meta)stability. 
The figure shows that the constraints from Br($\mu \rightarrow e \gamma$)
can sometimes further constrain the value of $y_\nu$ as obtained from 
vacuum metastability.  For instance, for $M_R = 200$ GeV   
and NH, the constraint from Br($\mu \rightarrow e \gamma$)
restricts $y_\nu$ to be $\leq 0.07$ for values of CP phases 
$(\delta, \alpha) = \left(0,\frac{3\,\pi}{2}\right)$. For $(\delta,\alpha) = (0,\frac{\pi}{2})$ 
the maximum allowed value of $y_\nu$ is lower. 
For other combinations of CP phases the bands lie anywhere inside or between the
two shaded regions. Thus if we consider all possible values of 
CP phases then only the region marked disallowed is not compatible with
the constraints from Br($\mu \rightarrow e \gamma$) for NH though it was 
consistent with vacuum metastability constraints. The vacuum stability constraints,
on the other hand, are stronger than those obtained from Br($\mu \rightarrow e \gamma$).

For IH and $(\delta,\alpha) = \left(0,\frac{3\,\pi}{4}\right)$, the hatched region 
extends all the way up to $M_R = 200$ GeV  and 
there is no significant constraint from $\mu \rightarrow e \gamma$ 
given the present uncertainty on the neutrino oscillation 
parameters. Nevertheless for the green hatched region 
corresponding to  $(\delta, \alpha) = (0,0)$,
the region to its left is disfavored and $y_\nu^{max}$ is constrained 
to lower values as compared to 
the bound from vacuum (meta)stability up to a certain value of $M_R$.  However, 
if we consider all possible values of CP phases then we can conclude that, 
given the present uncertainty of oscillation parameters and the CP phases, the 
vacuum (meta) stability bound on $(y_\nu)_{max}$ is stronger that that obtained from  
Br($\mu \rightarrow e \gamma$) for IH. 

We note in passing that in this type of models the Higgs boson can decay to two neutrinos
of which one is heavy and the other one is a light neutrino, as long as the heavy neutrino
is lighter than the Higgs boson. This has been studied in the context of inverse seesaw models and
put constraints on the Yukawa coupling $y_\nu$ to be $y_\nu \ls$ 0.02 for $M_R \ls$ 120 GeV from
the experimental data on the channel $h \rightarrow W W^* \rightarrow \ell \ell \nu \nu$ \cite{Dev:2012zg,DeRomeri:2012qd}.
For larger masses of the heavy neutrino current Higgs searches do not provide any constraint on the parameter space.

On the other hand, the search for heavy singlet neutrinos at LEP by the L3 collaboration in the decay channel
$N \rightarrow e W$ showed no evidence of such a singlet neutrino in the mass range between 80 GeV ($\left|V_{\alpha i}\right|^2\ls 2\times 10^{-3}$) and 205 GeV ($\left|V_{\alpha i}\right|^2\ls 1$) 
\cite{Achard:2001qv}. $V_{\alpha i}$ is the mixing parameter between the
heavy and light neutrino. Heavy singlet neutrinos in the mass range from 3 GeV up to the Z-boson mass ($m_Z$) has also
been excluded by LEP experiments from Z-boson decay up to $\left|V_{\alpha i}\right|^2\sim 10^{-5}$ \cite{Akrawy:1990zq,Adriani:1992pq,Abreu:1996pa}.
In the light of these experimental observations we have chosen the parameter $M_R$ to be greater than or equal to
200 GeV in this study.

\section{0$\nu \beta \beta$ decay}

The half life for neutrino-less double beta decay in 
presence of heavy singlets is given by \cite{0nunubb,Chakrabortty:2012mh},
\begin{eqnarray}
T_{(1/2)}^{-1}=
{G}
\frac{|\mathcal{M}_{\nu }|^2}{m_e^2}
\left| U^2_{L_{e\,i}}\, m_i +
<p^2> \frac{V^2_{e\,i}}{M_i}
\right|^2,
\label{thalf} 
\end{eqnarray}
where $<p^2>$ is given by \cite{Tello:2010am}
\begin{eqnarray}
<p^2>=-m_e m_p\frac{\mathcal{M}_N}{\mathcal{M}_{\nu}}.
\end{eqnarray}
$\mathcal{M}_{\nu}$ and $\mathcal{M}_{N}$
denote the nuclear matrix elements corresponding to light and heavy neutrino
exchange respectively.
The values of the parameters are taken as \cite{0nunubb}
$G=7.93\times10^{-15}$ yr$^{-1}$, 
$<p^2>=-(182 \text{ MeV})^2$.

The first term in Eq. (\ref{thalf}) is the usual contribution from the 
left-handed neutrinos. The second term denotes the contribution of 
the singlet neutrinos.  
The matrix V is defined in Eq. (\ref{bdmatrix}).  
Taking the most general form of the matrix $m_D^\prime$ as
\begin{equation} 
m_D^\prime = 
\begin{pmatrix} 
m_{d1}  & m_{d2} & m_{d3} \\
m_{s1} & m_{s2} &  m_{s3}
\end{pmatrix}, 
\end{equation} 
and $U_R$ and $M$ as defined in Eq. (\ref{ur}) and (\ref{msinglet}) 
respectively,  we obtain, 
\begin{equation} 
V_{e1} = \frac{i}{\sqrt{2}~M_R} (m_{s1}^{\ast} - m_{d1}^{\ast}),
V_{e2} = \frac{1}{\sqrt{2}~M_R}(m_{s1}^{\ast} +  m_{d1}^{\ast}). 
\end{equation} 
Then the contribution from the heavy part  is
$ 2 <p^2> m_{s1}^{\ast} m_{d1}^{\ast}/M_R^3 \sim 10^{-8} m_i$. 
Thus this contribution is negligible as compared to the 
contribution from the light sector which is $\sim m_i$. 
\begin{figure}[ht]
\begin{center}
\includegraphics[width=8.0cm,height=5.5cm]{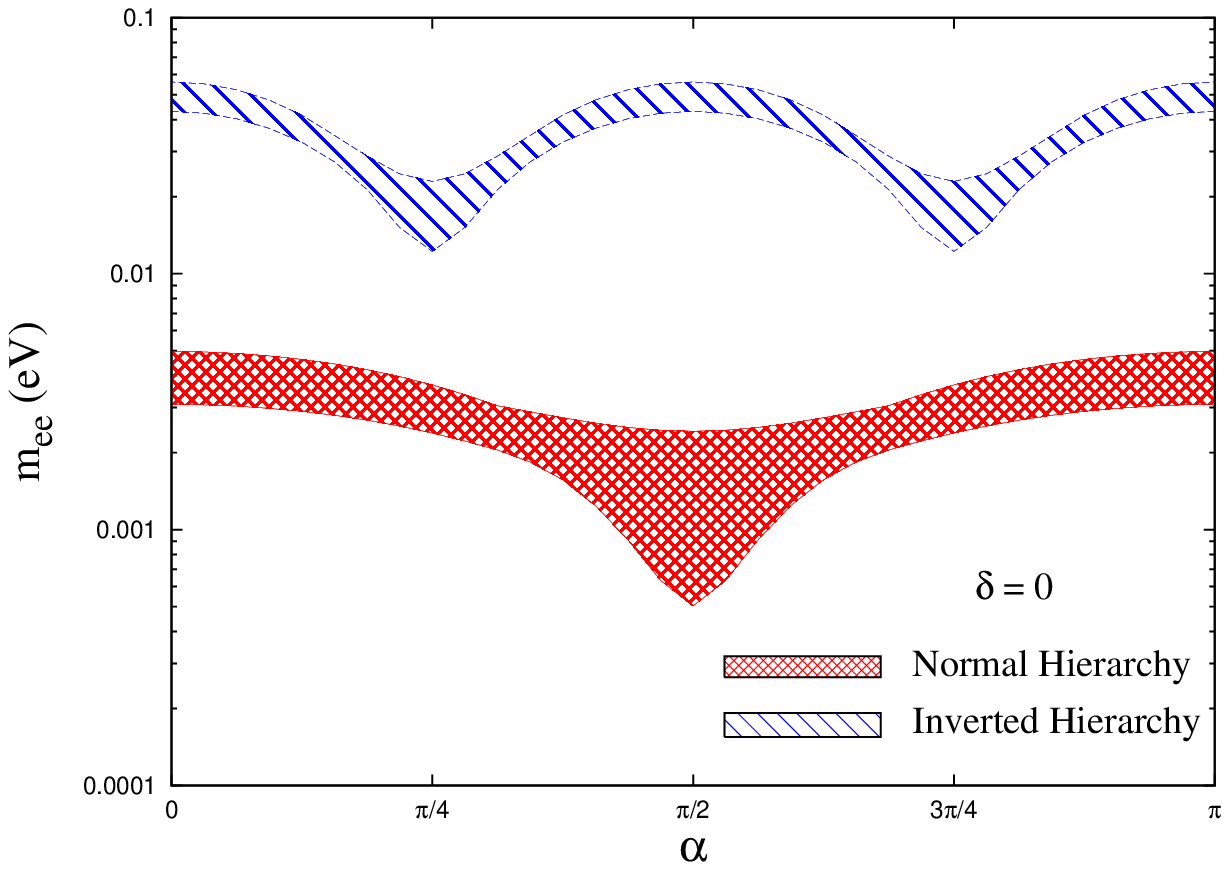}
\includegraphics[width=8.0cm,height=5.5cm]{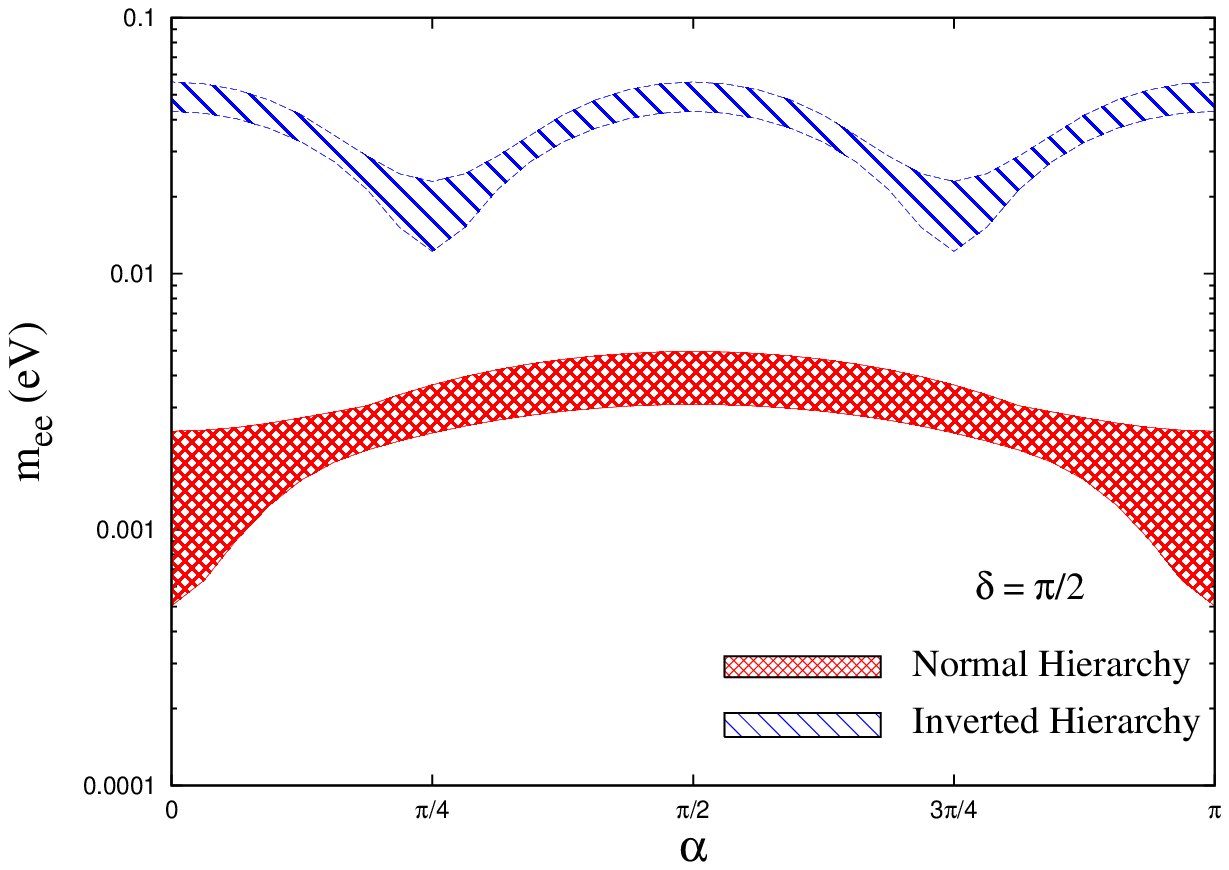}
\caption{The effective mass governing $0\nu\beta\beta$ as a function of
the Majorana phase $\alpha$ for  NH (dark (red) shaded curve) and IH (light (blue)
shaded curve). The left panel is for $\delta=0$ while the right panel
is for $\delta=\pi/2$.
}
\label{fig:0nubb}
\end{center}
\end{figure}
 
Therefore, $0\nu\beta\beta$ is due to the light neutrinos 
only and the effective mass is defined as 
\begin{equation} 
m_{ee} = \left| U^2_{L_{e\,i}}\, m_i  \right|.
\end{equation}
Since in this case the lightest mass is zero one can plot the 
conventional plots of effective mass as a function of the unknown 
CP phases for both hierarchies. 

For NH the effective mass $m_{ee}$ in the limit of the
smallest mass $m_1 \rightarrow 0$ 
is given as,  
\begin{equation} 
|m_{ee}|_{NH} = \sqrt{\Delta m^2_{atm}}~  
\left|\sqrt{r} s_{12}^2 c_{13}^2 e^{ 2 i \alpha} + s_{13}^2 e^{-2 i \delta} \right|. 
\label{mnuee_xnh} 
\end{equation}
The maximum is obtained for $(\alpha, \delta) = (0,0)$ or $(\pi/2,\pi/2)$
while the minimum occurs for $(\alpha, \delta) = (0,\pi/2)$ or $(\pi/2,0)$. 
This is reflected in Fig. \ref{fig:0nubb} by the dark (red) shaded 
curve which represents the effective mass governing $0\nu\beta\beta$ 
as a function of the Majorana phase $\alpha$.   
The shaded portion is due to the 3$\sigma$ uncertainty in the 
oscillation parameters that appear in the  expression of effective mass.
The left panel is for $\delta =0$ and the right panel is for 
$\delta = \pi/2$.  
The cancellation condition is 
\be
\sqrt{r}~sin^ 2\theta_{12} = \tan^2\theta_{13},
\ee
which is not satisfied for the current 3$\sigma$ ranges of parameters
and therefore the effective mass does not vanish, which is 
also seen from the figure. 
For IH the smallest mass is $m_3$ which is zero in this model and the 
effective mass is 
\begin{equation}
\left|m_{ee}\right|_{IH} =\sqrt{\Delta m^2_{atm}}(c_{12}^2 c_{13}^2 e^{- 2 i \alpha }  + s_{12}^2 c_{13}^2 e^{2i\alpha}). 
\end{equation}
For IH the effective mass is independent of the Dirac phase $\delta$. 
The maximum of 
$\left|m_{ee}\right|$ occurs for $\alpha = 0, \pi/2,\pi$  
and the corresponding expression is,
\be
\left|m_{ee}\right|_{max}=c_{13}^2\sqrt{\Delta m^2_{atm}}.
\label{meeihmax}
\ee
The minimum value is obtained for $\mathrm{\alpha=\pi/4, 3 \pi/4}$ as,
\be
\left|m_{ee}\right|_{min}= c_{13}^2 cos\,2\theta_{12}\sqrt{\Delta m^2_{atm}}.
\label{meeihmin}
\ee
This  is seen from  Fig. \ref{fig:0nubb} by the light (blue) shaded curve. 
$m_{ee}$ for IH is in the range accessible to future neutrinoless double 
beta decay experiments.

\section{ Collider Signatures} 

As mentioned earlier, if the heavy singlet neutrinos
have mass less than the Higgs boson mass, then the Higgs boson
can have new decay modes \cite{Dev:2012zg}. For example, the Higgs boson can decay into $h \rightarrow {\bar \nu} N$.
Now, the singlet neutrinos can decay into $l W$ and $\nu Z$ through the mixing between the heavy neutrinos 
and the light active neutrinos. At the LHC this will lead to final states such
as $pp \rightarrow h \rightarrow \ell^+ \ell^- + \mET$, where $\ell = e, \mu$. Note that these final states will depend on 
the Yukawa couplings and one can put bounds on these Yukawa couplings from the existing LHC data on these types of final states
\cite{Dev:2012zg}. 

We have considered the singlet neutrino to be heavier than the Higgs boson. In this case one has to look at the 3-body decay 
modes of the Higgs boson through the virtual heavy neutrino to have similar final states. Obviously, in this case the constraints
on the Yukawa couplings will be much less restrictive. In our model we have obtained upper bound on the Yukawa couplings $y_\nu$ 
from the vacuum stability condition and this can be used to test our model at the LHC by looking at the dilepton plus missing $E_T$ final states. 

One can also have trilepton plus missing $E_T$ final states at the LHC from the production of these heavy neutrinos
\cite{Das:2012ze}. For example, at the LHC the heavy neutrinos can be produced through the s-channel $W^\pm$ exchange: $u {\bar d}
\rightarrow \ell^+ N$ or $u {\bar d} \rightarrow \ell^+ S$. $N$ or $S$ can again decay into $l W$ and $\nu Z$ through $\nu-N$ or $\nu-S$
mixing. This will lead to trilepton plus missing $E_T$ final states at the LHC. Now, the trilepton plus $\mET$ signal is a very clean
signal for looking at physics beyond the standard model. In our model, the trilepton final states depend once again on the Yukawa couplings
of our model. Using the upper bound on $y_\nu$ obtained from vacuum (meta)stability condition, it would be possible to study the present model at the
LHC through the trilepton channel. Note that in this model the lepton number violating
like-sign di-lepton signal will be suppressed at the colliders
because of smallness of $y_s$.

\section{Conclusion}

In this paper we consider the phenomenology of
the minimal linear seesaw model
consisting of three left-handed neutrinos and two singlet 
fields. The two singlet fields have opposite lepton numbers.  
Smallness of neutrino mass is ensured in this model by the tiny lepton number 
violating coupling ($y_s$) of
one of the  singlets  with the left-handed neutrinos.
Thus, the masses ($M_R$) of the heavy singlet neutrinos can be at the
TeV scale  even with the Dirac type 
coupling ($y_\nu$) between the other singlet and the heavy state 
of ${\cal{O}}(1)$.  
This permits appreciable light-heavy mixing in the model 
which can have interesting phenomenological consequences.  
The model predicts one massless neutrino and hence there is only
one Majorana phase. The great advantage of this model is that 
the Yukawa matrices can be fully reconstructed            
in terms of the oscillation parameters apart from the overall coupling
strengths $y_\nu$ and $y_s$. 

The presence of the coupling $y_\nu$ in seesaw models tends to destabilize
the vacuum further as compared to the SM. 
This allows one to constrain the coupling strength $y_\nu$ from consideration of
vacuum (meta)stability. Since the absolute stability of the electroweak vacuum is 
already severely restricted by the present value of the Higgs mass, this puts 
a very stringent constraint: $y_\nu \ls 0.07$ in a limited region of the
parameter space. 
On the other hand, we show that consideration of the  metastability 
allows one to constrain the coupling strength $y_\nu$ as
$y_\nu \ls$ 0.48 for 
$200~{\rm GeV}~ \ls ~M_R \ls 1 ~{\rm TeV}$. Both these
bounds depend on the value of the strong coupling constant ($\alpha_s$),
the top quark mass ($m_t$) and the Higgs boson mass ($m_h$). 
Our analysis includes the effect of neutrino correction to the effective potential
at the one loop level. 

Bounds on $y_\nu/M_R$ can be obtained as a function of the CP phases 
$\alpha$ and $\delta$ 
from experimental constraints on  
lepton flavor violating processes. 
Combined constraints from vacuum (meta)stability and 
the lepton flavor violating process $\mu \rightarrow e \gamma$ 
rule out a significant portion of
the parameter space in the ($y_\nu$--$M_R$) plane  for NH  
for masses of $M_R \ls (200-2000)$ GeV depending on the values of the other parameters. 
On the other hand, contribution of the singlet
neutrinos to the neutrinoless double beta decay process is insignificant.
The model predicts interesting signatures at the LHC and can be tested using 
the present and future data. 
However, a complete collider study  merits a separate analysis.

\section{Acknowledgments} The authors wish to thank the sponsors and 
organizers of the Workshop on High Energy Physics Phenomenology (WHEPP-XI), 
where the work on this problem started.   We  want to thank  
J. Chakrabortty, D. Ghosh, A. Ibarra
A. Joshipura, N. Mahajan, S. Mohanty, I. Saha for helpful discussions. 
S.K. wishes to acknowledge discussions with S. Acharyya, D. Angom,
V. Kishore, S. Raut regarding computational analysis.  
S.K. and S.G wishes to acknowledge the hospitality at the 
Department of Theoretical Physics, 
Indian Association for the Cultivation of Science
during the course of this work.

\end{document}